\documentclass[preprint2]{aastex}
\usepackage{url}
\usepackage{natbib}
\usepackage{color}

\newcommand{\HI}{{\sc H}\,{\scriptsize{\sc I}}}

\shorttitle{CR gradient in the 3rd quadrant with \textit{Fermi}}
\shortauthors{Ackermann et al.}

\begin{document}

\title{
Constraints on the Cosmic-Ray Density Gradient beyond the Solar Circle from
\textit{Fermi} $\gamma$-ray Observations of the Third Galactic Quadrant
}

\author{
\small
M.~Ackermann\altaffilmark{2}, 
M.~Ajello\altaffilmark{2}, 
L.~Baldini\altaffilmark{3}, 
J.~Ballet\altaffilmark{4}, 
G.~Barbiellini\altaffilmark{5,6}, 
D.~Bastieri\altaffilmark{7,8}, 
K.~Bechtol\altaffilmark{2}, 
R.~Bellazzini\altaffilmark{3}, 
B.~Berenji\altaffilmark{2}, 
E.~D.~Bloom\altaffilmark{2}
% ...
}
% ...
\altaffiltext{2}{W. W. Hansen Experimental Physics Laboratory, Kavli Institute for Particle Astrophysics and Cosmology, Department of Physics and SLAC National Accelerator Laboratory, Stanford University, Stanford, CA 94305, USA}
\altaffiltext{3}{Istituto Nazionale di Fisica Nucleare, Sezione di Pisa, I-56127 Pisa, Italy}
\altaffiltext{4}{Laboratoire AIM, CEA-IRFU/CNRS/Universit\'e Paris Diderot, Service d'Astrophysique, CEA Saclay, 91191 Gif sur Yvette, France}
\altaffiltext{5}{Istituto Nazionale di Fisica Nucleare, Sezione di Trieste, I-34127 Trieste, Italy}
\altaffiltext{6}{Dipartimento di Fisica, Universit\`a di Trieste, I-34127 Trieste, Italy}
\altaffiltext{7}{Istituto Nazionale di Fisica Nucleare, Sezione di Padova, I-35131 Padova, Italy}
\altaffiltext{8}{Dipartimento di Fisica ``G. Galilei", Universit\`a di Padova, I-35131 Padova, Italy}
% ...

\begin{abstract}

We report an analysis of the interstellar $\gamma$-ray emission
in the third Galactic quadrant measured by the {Fermi} Large Area Telescope.
The window encompassing the Galactic plane from longitude 
$210\arcdeg$ to $250\arcdeg$ has
kinematically well-defined segments of the Local and the Perseus arms, suitable to study 
the cosmic-ray densities across the outer Galaxy. 
We measure no large gradient with Galactocentric distance 
of the $\gamma$-ray emissivities
per interstellar H atom over the regions sampled in this study. The
gradient depends, however,  on the optical depth correction applied to derive the 
\HI\ column densities.
No significant variations are found in the interstellar spectra in the outer Galaxy, indicating
similar shapes of the cosmic-ray spectrum up to the Perseus arm for particles with GeV to tens of
GeV energies.
The emissivity as a function of Galactocentric radius does not show a large enhancement 
in the spiral arms with respect to the interarm region.
The measured emissivity gradient is flatter than expectations based on a cosmic-ray propagation model using the radial 
distribution of supernova remnants and uniform diffusion properties. In this context, observations
require a larger halo size and/or a flatter CR source distribution than usually assumed.
The molecular mass calibrating ratio, 
$X_{\rm CO} = N({\rm H_{2}})/W_{\rm CO}$, 
is found to be $(2.08 \pm 0.11) \times 10^{20}~{\rm cm^{-2}~(K~km~s^{-1})^{-1}}$ in the Local-arm
clouds
and is not significantly sensitive to the choice of \HI\ spin temperature. No significant variations
are found for clouds in the interarm region.
\end{abstract}


\section{Introduction}

Knowledge of the distribution of cosmic-ray (CR) densities within our Galaxy is a key
to understanding their origin and propagation.
High-energy CRs interact with the gas in the interstellar medium (ISM)
or the interstellar radiation field,
and produce $\gamma$-rays via nucleon-nucleon interactions,
electron Bremsstrahlung and inverse Compton (IC) scattering.
Since the ISM is transparent to these $\gamma$-rays, we can
probe CRs in the local ISM, beyond direct measurements performed in the solar system, as well
as in remote locations of the Galaxy. Although much effort has been made since the COS-B era
\citep[e.g.,][]{Strong1988,Strong1996}, the results have been limited by
the angular resolution, effective area and energy coverage of the instruments.
The advent of the \textit{Fermi} Gamma-ray Space Telescope
enables studying the spectral and spatial distribution of diffuse
$\gamma$-rays and CRs with unprecedented sensitivity.

Here we report an analysis of diffuse $\gamma$-ray emission observed
in the third Galactic quadrant. The window with Galactic longitude  
$210\arcdeg \le l \le 250\arcdeg$ and latitude $-15\arcdeg \le b \le +20\arcdeg$ 
hosts kinematically well-defined segments of the Local and the Perseus spiral arms
and is one of the best regions to study the CR density distribution across the outer Galaxy.
The region has been already studied by  \citet{Digel2001} using EGRET data.
The improved sensitivity and angular resolution of the \textit{Fermi} LAT \citep[Large Area Telescope;][]{Atwood2009}
and recent developments in the study of the ISM allow us to examine the CR spectra and density distribution with better accuracy.
We exclude from the analysis the region 
of the Monoceros R2 giant molecular cloud and the Southern Filament
of the Orion-Monoceros complex \citep[e.g.,][]{Wilson2005},
in $l \leq 222\arcdeg$ and $b \leq -6\arcdeg$, because 
1) star forming activity and possible high magnetic fields
suggested by the filamentary structure
\citep[e.g.,][]{Morris1980,Maddalena1986}
could indicate a special CR environment, and
2) an OB association in Monoceros R2 may hamper
the determination of ISM densities from dust tracers
(see \S~2.1.2 for details).

Study of the $X_{\rm CO}$ conversion factor which transforms the integrated intensity of the 2.6~mm
line of carbon monoxide, $W_{\rm CO}$, into the molecular hydrogen column density, $N({\rm H_{2}})$,
is also possible since the region contains well-known molecular complexes. In the Local arm we find
the molecular clouds 
associated with Canis Major OB~1, NGC~2348 and NGC~2632 
\citep{Mel'nik1995,Kaltcheva2000}. At a few kpc from the Solar System, in the interarm, lower-density region
located between the Local and Perseus arms, we find Maddalena's cloud
\citep{Maddalena1985}, a giant molecular cloud remarkable for its lack of
star formation, and the cloud associated with Canis Major OB~2 \citep{Kaltcheva2000}.

This study complements the \textit{Fermi} LAT
study of the Cassiopeia and Cepheus region in the
second quadrant reported by \citet{Abdo2010_2nd}.
The paper is organized as follows. We describe the model preparation in \S~2
and the $\gamma$-ray observations, data selection and the analysis procedure in \S~3.
The results are presented in \S~4, where we also discuss the emissivity profile measured for the
atomic gas
and we compare it with predictions by a CR propagation model.
A summary of the study is given in \S~5.

\section{Modeling the Gamma-Ray Emission}

\subsection{Interstellar Gas}

\subsubsection{\HI\ and {\sc CO}}\label{hicosec}

In order to derive the $\gamma$-ray emissivities associated with the different components of the ISM
we need to determine the interstellar gas column densities separately for each region and gas
phase.
For atomic hydrogen we used the Leiden/Argentine/Bonn Galactic \HI\ survey 
by \citet{Kalberla2005}. In order to turn the \HI\ line intensities into $N$(\HI) column densities, 
a uniform spin temperature $T_{\rm S}=125~{\rm K}$ has often been adopted in previous studies. 
We will consider this option to directly
compare our results with the former EGRET analysis of the same region \citep{Digel2001} 
and other studies of the
Galactic diffuse emission by the LAT \citep{Abdo2009_HI,Abdo2010_2nd}. Recent \HI\ absorption
studies \citep{Dickey2009}, however, point to larger average spin temperatures 
in the outer Galaxy, so we have tried
different choices of $T_{\rm S}$ to evaluate how the optical depth correction affects the results.
We will find that the emissivity per \HI\ atom and the inferred CR density
is affected by up to $\sim$~50\% in the Perseus arm, and will take this uncertainty into account in the discussion.

The integrated intensities of the 2.6~mm line of CO, 
$W_{\rm CO}$, have been derived from the composite survey by \citet{Dame2001}.
The data have been filtered with the moment-masking technique in order to reduce the noise 
while keeping the resolution of the original data.

\begin{figure}
\includegraphics[width=0.5\textwidth]{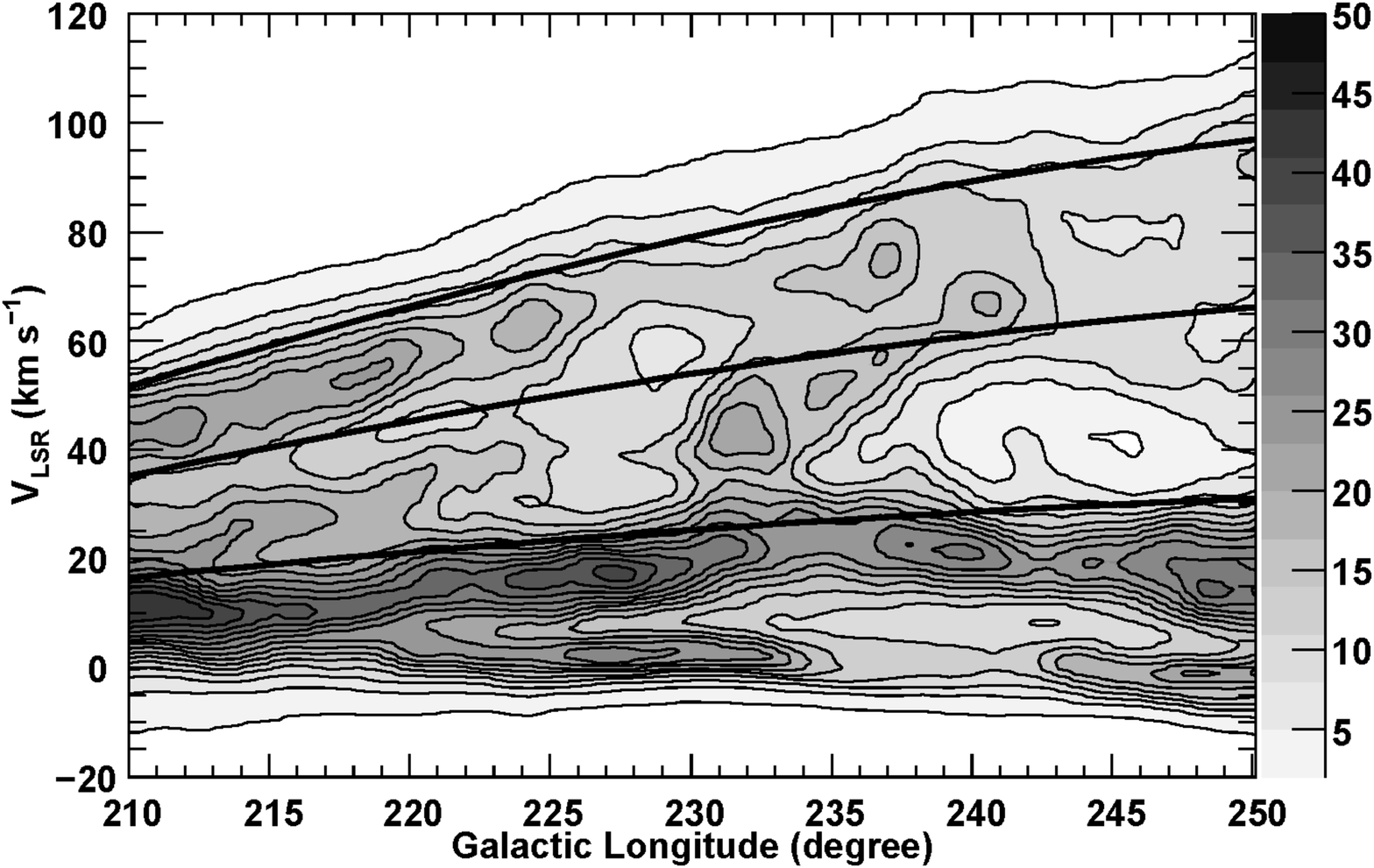}
\caption{
Longitude-velocity diagram of the average intensity 
of the 21~cm line (in unit of K) for $-15\arcdeg \le b \le 20\arcdeg$.
Preliminary boundaries between the four Galactocentric annuli
are also presented (See \S~2.1.1 for details).
The lowest contour corresponds to 2~K and the contour interval is 3~K.
}
\label{fig:Diagram}
\end{figure}

Figure~\ref{fig:Diagram} shows the velocity-longitude profile
of \HI\ emission in our region of interest (ROI).
The preparation of maps accounting for the different Galactic structures
present along the line of sight 
is similar to that described in detail
in \citet{Abdo2010_2nd} and based on a sequence of three steps:
\begin{enumerate}
\item preliminary separation within Galactocentric rings;
\item adjustment of the boundaries based on the velocity structures of the
interstellar complexes;
\item correction for the spill-over due to the velocity dispersion of the
broad \HI\ lines between adjacent regions.
\end{enumerate}

Four regions were defined in Galactocentric distance, namely
the Local arm (Galactocentric radius $R \le {\rm 10~kpc}$), the
interarm region ($R=10\mbox{--}12.5~{\rm kpc}$), the Perseus arm
($R=12.5\mbox{--}16~{\rm kpc}$) and the region beyond the Perseus arm (which hosts a faint segment
of the outer arm; $R\ge{\rm 16~kpc}$).
The boundaries separating these regions
under the assumption of a flat rotation curve \citep{Clemens1985} for the case
of
$R_{0} = 8.5~{\rm kpc}$ and 
$\theta_{0} = 220~{\rm km~s^{-1}}$
(where $R_{0}$ and $\theta_{0}$ are the Galactocentric radius
and the orbital velocity of the local group of stars, respectively)
are overlaid in Figure~\ref{fig:Diagram}.

The preparation of the \HI\ and CO gas maps started from these preliminary
velocity boundaries, which were then adjusted
for each line of sight to the closest minimum in the \HI\ spectrum
\footnote{
\ The minima are
unlikely to be due to self absorption, because the velocity-distance relation is single valued
in the outer Galaxy.}.
Then, the spill-over from one velocity interval to the next ones due to the velocity dispersion for
the broad \HI\ lines was corrected by fitting each \HI\ spectrum with a combination of Gaussian
profiles. 
We believe that this separation procedure provides more
accurate estimates of the ISM column densities of each Galactic region than a simple slicing based
on the rotation curve. 

In particular, effort was put into separating the outer arm structures from the more massive Perseus arm component, 
especially at $l \gtrsim 235\arcdeg$ where the \HI\ lines from the two regions 
merge into a single broad component. 
For directions where a minimum in the \HI\ brightness temperature profile was not found near
the $R = 16$ kpc velocity boundary, 
we integrated the profiles on both sides of the $R = 16$ kpc velocity boundary 
to estimate the Perseus and outer arm contributions. Then we inserted a line in the Gaussian fitting
at the outer-arm velocity extrapolated from the $l-v$ trend 
observed at $l \gtrsim 235\arcdeg$ to correct for the spill-over due to the velocity dispersion.
Given these difficulties we expect large systematic uncertainties in the outer-arm $N$(\HI) column
densities and the
corresponding $\gamma$-ray emissivities will not be
considered for the scientific interpretation.
We note that the impact on the emissivities associated with the
inner regions is small, $\le 10$\% as described in \S~\ref{sec:CRgradient}.

\begin{figure}
\includegraphics[width=0.5\textwidth]{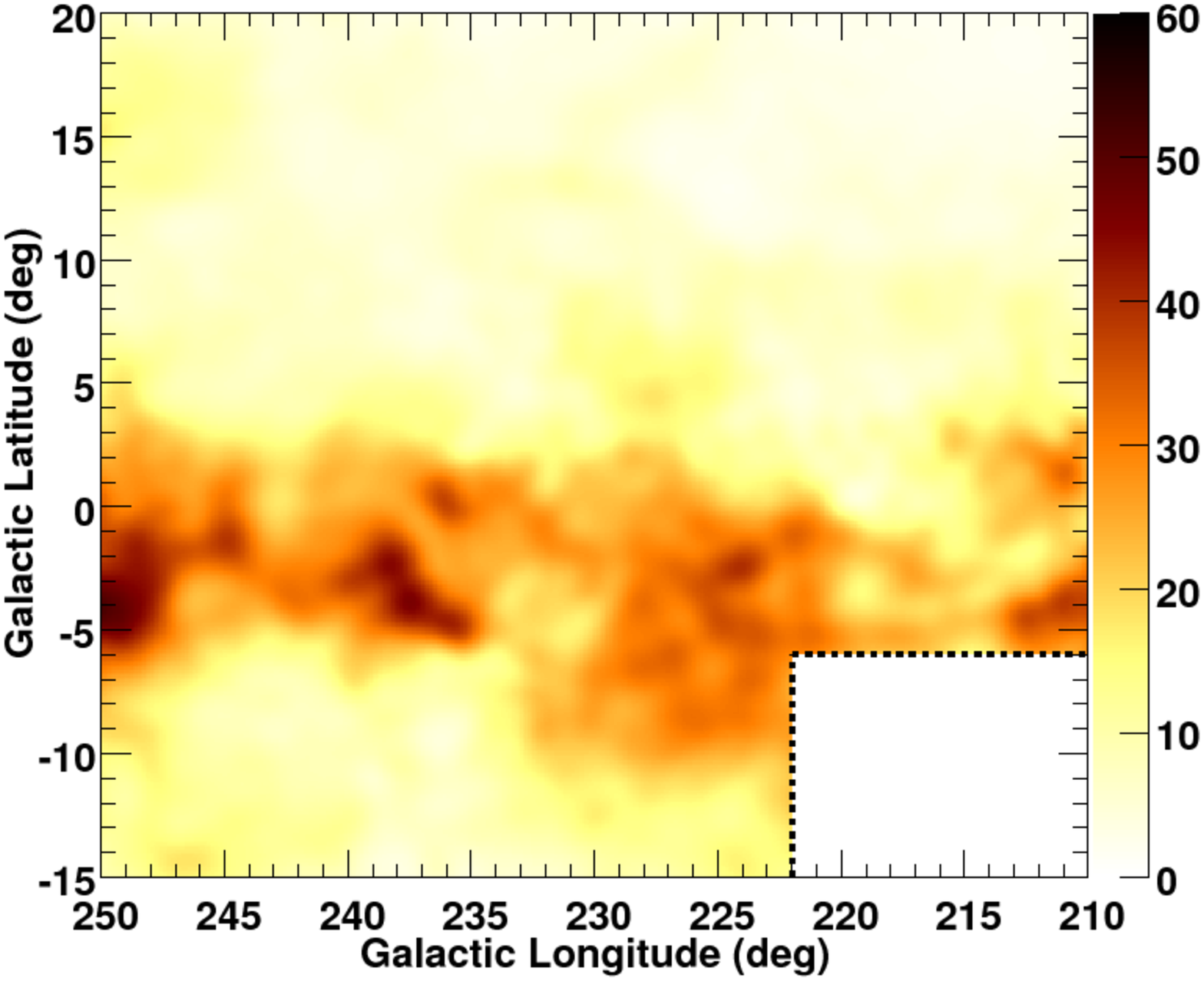}
\includegraphics[width=0.5\textwidth]{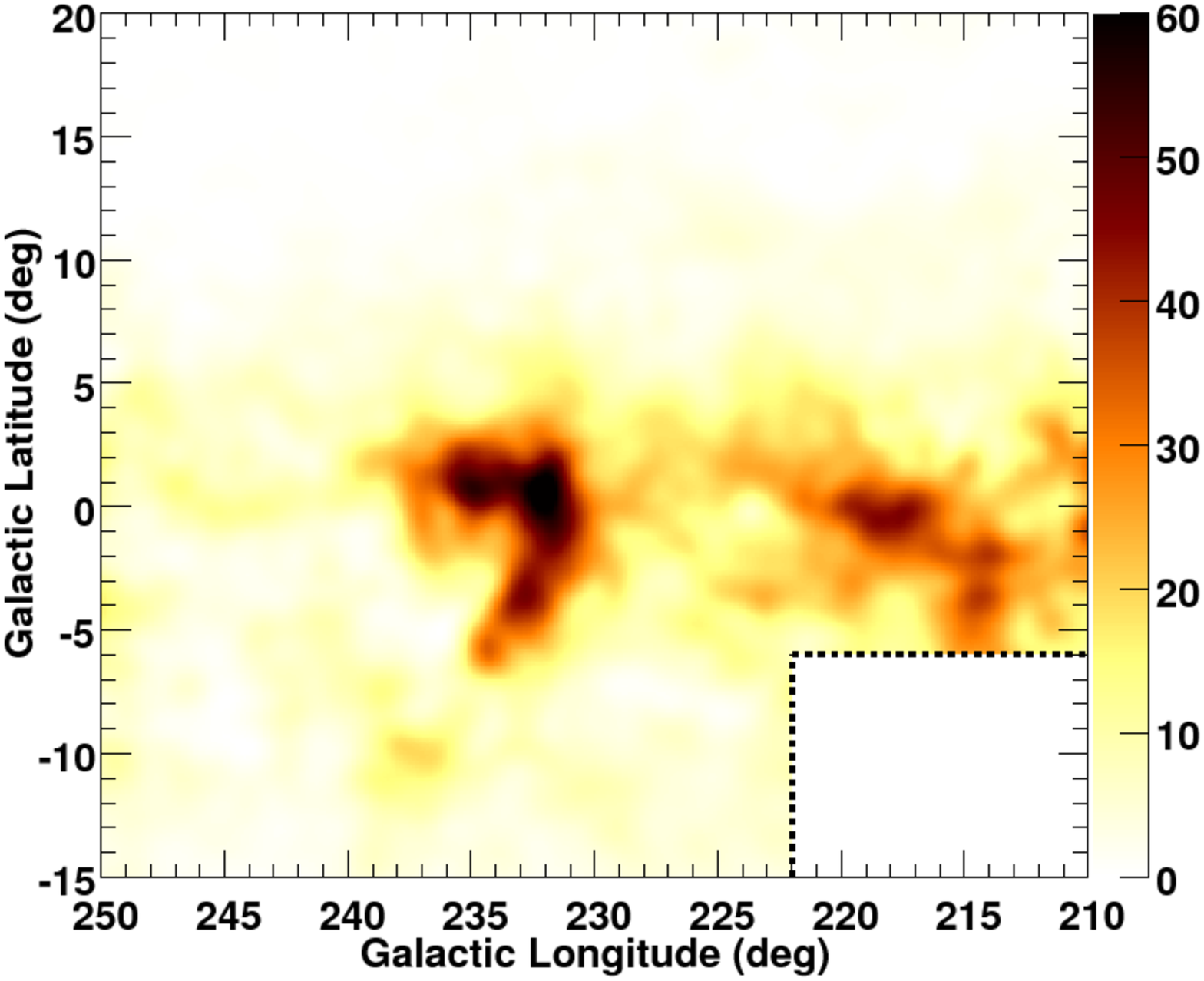}
\includegraphics[width=0.5\textwidth]{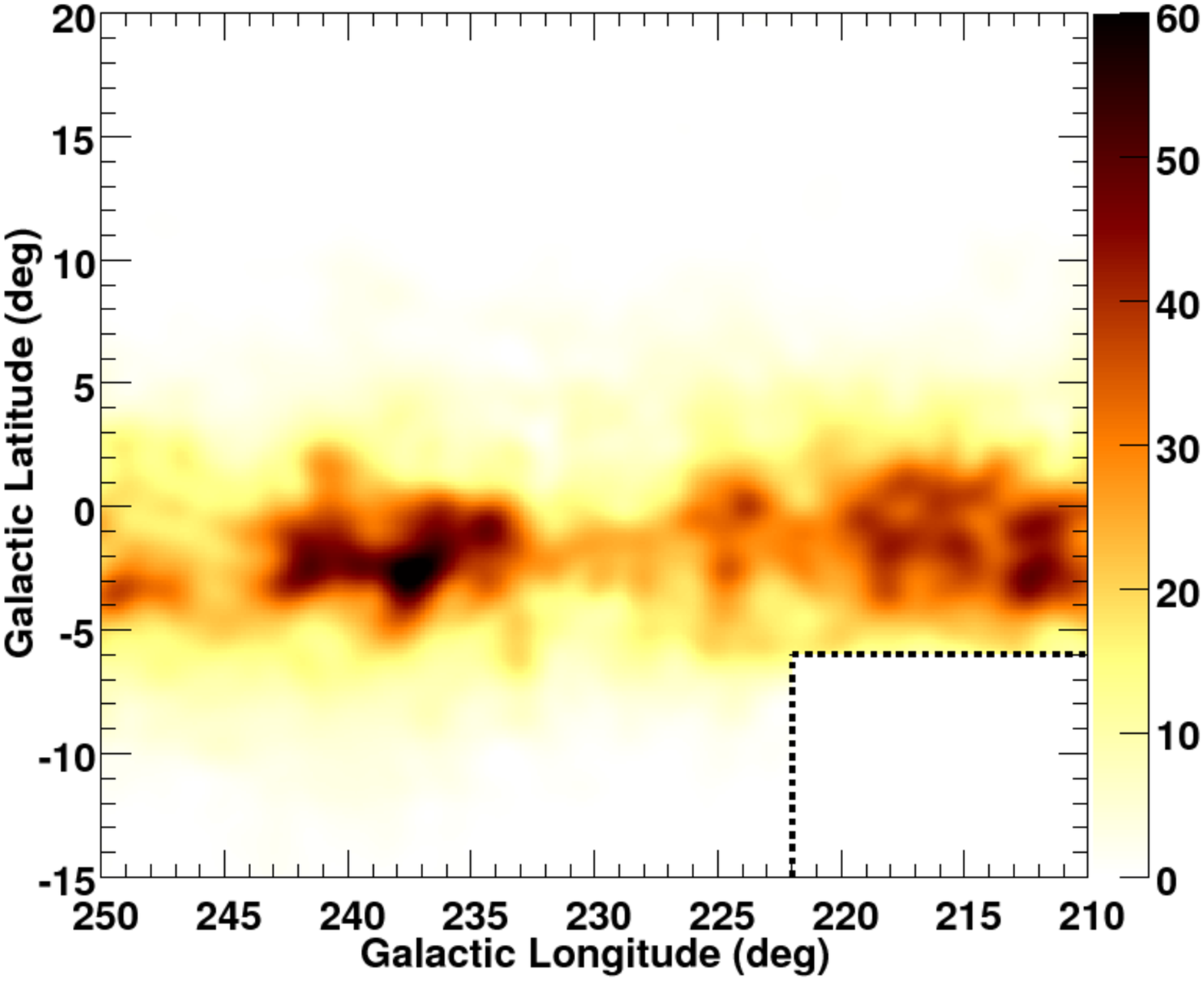}
\includegraphics[width=0.5\textwidth]{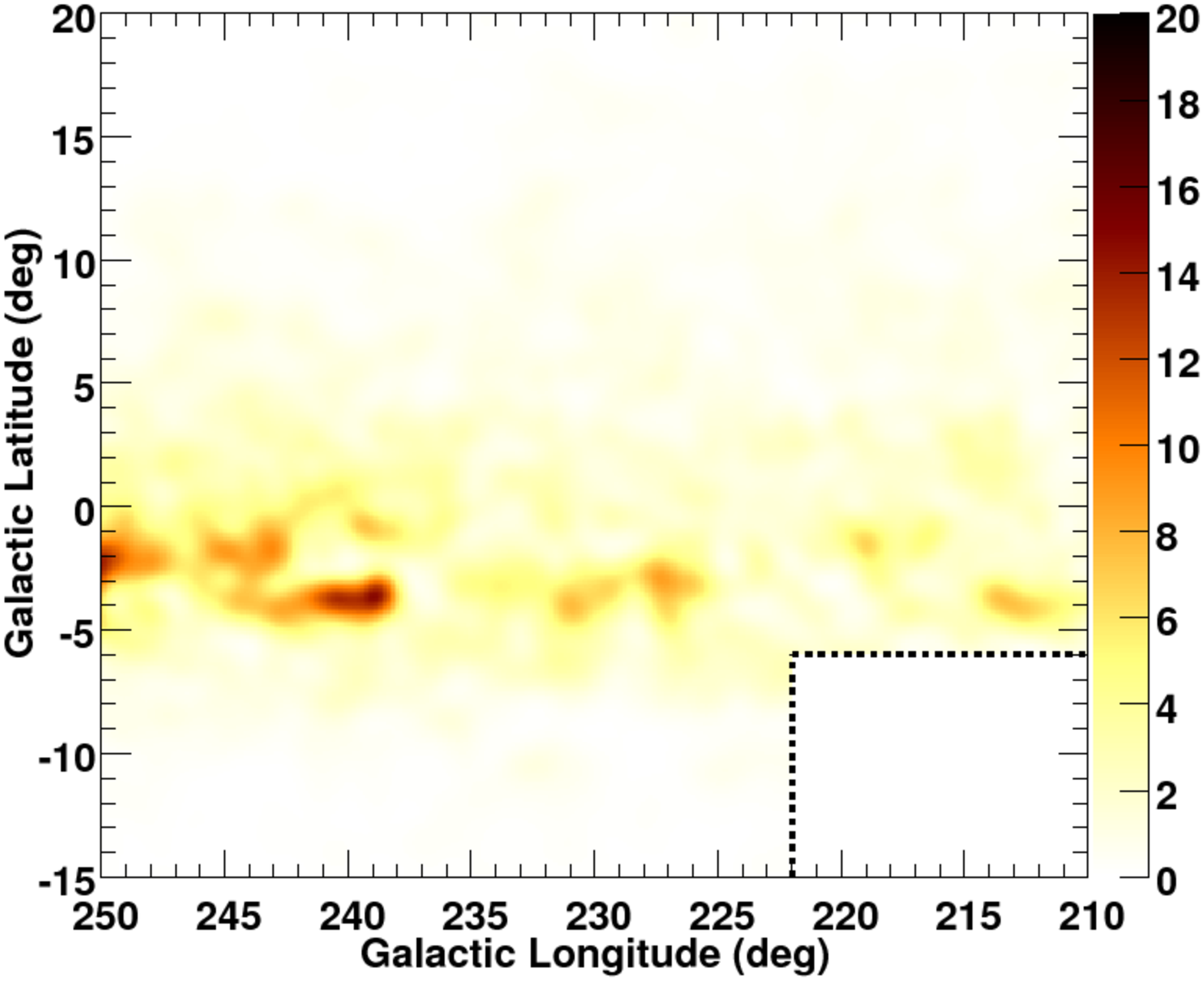}
\caption{
Maps of $N$(\HI) (in unit of ${\rm 10^{20}~atoms~cm^{-2}}$)
for the Local arm (top left), interarm (top right), Perseus arm (bottom left), and outer arm (bottom
right) regions, obtained for a spin temperature
$T_{\rm S}=125~{\rm K}$.
The outlined area in the bottom right corner is not used
in the analysis (see \S~1).
The maps have been smoothed with a Gaussian with $\sigma = 1\arcdeg$ for display.
}
\label{fig:HI}
\end{figure}

The resulting maps are shown in Figure~\ref{fig:HI} and \ref{fig:CO}.
They exhibit a low level of spatial degeneracy, and thus allow us to
separate the $\gamma$-radiation arising from the interaction with CRs in each component.

\subsubsection{Interstellar reddening}

It has been long debated whether the combination of \HI\ and CO surveys traces total column densities of neutral interstellar matter.
By comparing gas line surveys, the $\gamma$-ray observations by EGRET and dust thermal emission, \citet{Grenier2005}
reported a considerable amount of neutral gas at the interface between the two \HI\ and CO emitting
phases, associated
with cold dust but not properly traced by \HI\ and CO observations.
Their finding was then confirmed by LAT data for the Gould Belt in the second quadrant
\citep[][]{Abdo2010_2nd}.

In order to complement the \HI\ and CO maps, we have prepared a map derived from the
$E(B-V)$ reddening map by \citet{Schlegel1998}. The residual point sources at low latitudes were masked by
setting to zero regions of $0.2\arcdeg$ radius centered on the
positions of potential
IRAS point sources\footnote{\
\url{http://cdsarc.u-strasbg.fr/viz-bin/Cat?II/274}.
See \citet{IRAScat}.} if the $E(B-V)$ magnitude exceeded by $\gtrsim 20$\% 
that in surrounding pixels. 
The masked regions were then restored through an inpainting technique \citep{Elad2005}.
In the course of the work, various source masking techniques have been used with negligible
impact on the \HI\ and CO emissivity results.

The resulting map was fitted with a linear combination of the set of
$N$(\HI) and $W_{\rm CO}$ maps described in \S~2.1.1. The operation was repeated for different choices of
\HI\ spin temperature. The fit was performed over the same region as for the
$\gamma$-ray analysis, excluding a $3\arcdeg \times 3\arcdeg$ region centered
around Canis Major OB~1 \citep{Mel'nik1995} where the temperature correction applied by
\citet{Schlegel1998} to construct the $E(B-V)$ map from the dust thermal
emission is highly uncertain. A preliminary fit had led to extremely negative
residuals ($\le -1$ mag) around $l=224\arcdeg$, $b=-3\arcdeg$. 
Therefore, the residual $E(B-V)$ map was calculated masking this region in the fit. 
We are aware that the temperature corrections used by 
\citet{Schlegel1998} are less reliable with decreasing latitude, 
but the improvement we find in the $\gamma$-ray fit by adding the dust residual map supports the use of their map at low latitude.

The residual $E(B-V)_{\rm res}$ map, after subtracting the linear combination
of $N$(\HI) and $W_{\rm CO}$ maps, 
is shown in Figure~\ref{fig:EBV} (left panel).
The residuals typically range from $-0.5$ to $+0.5$ magnitude.
Large regions of positive residuals are found along the Galactic plane, in
association with molecular/atomic clouds. They are expected to trace gas
not correctly traced by \HI\ and CO surveys. A remarkable region of positive
residuals is detected at intermediate latitudes around $l=245\arcdeg$,
$b=+17\arcdeg$, in a region not covered by CO surveys. It corresponds to positive residuals also in $\gamma$-rays 
(\S~\ref{sec:procedure}) and may be due
to a missing, but possibly CO-bright molecular cloud (already suggested by \citealt{Dame2001} discussing
the completeness of their survey, see Figure~8 of their paper).
The negative residuals are generally small and may result from limitations in the gas column density
derivation and/or dust spectral variations.
The dust residual map compares well with the $\gamma$-ray residual map obtained when
using only \HI\ and CO to model the $\gamma$-ray emission (Figure~\ref{fig:EBV}, right panel). The
correlation between the spatial distributions of the dust and $\gamma$-ray residuals is
statistically confirmed in \S~\ref{sec:procedure}. Dust and $\gamma$-rays are consistent with the
presence
of missing gas
in the positive residual clumps. The faint ``glow'' of negative residuals on both sides of the
Galactic plane is driven by the nearby $N$(\HI) maps and it remains even when using the smallest
possible column-densities derived in the optically thin case. It may suggest a small change in
average spin temperature from the massive, compact clouds sampled in the plane to the more diffuse
envelopes sampled off the plane, or it may be due to the presence of more missing gas in the plane
than our templates can provide for in the fit. The dust-to-gas ratio as well as the $\gamma$-ray
emissivity in the \HI\ components would then be driven to higher values by the low latitude data and
would slightly overpredict the data off the plane.
	
The interpretation of the $E(B-V)_{\rm res}$ map in this region of the sky is complicated by the lack
of distance information for the dust emission. It is not possible to
unambiguously assign the residuals to any of the regions under study. Since we aim at separating different regions along the
lines of sight to investigate the CR density gradient in the outer Galaxy,
using the \HI\ and CO lines is essential. We have therefore used the $E(B-V)_{\rm res}$ map 
to correct for the total gas column densities. 
This approach is supported by the correlation we find between the dust and $\gamma$-ray data
(\S~\ref{sec:procedure}).
We also note that, since the dust contribution linearly correlated with 
the \HI\ and CO maps has been removed in the $E(B-V)_{\rm res}$ map, this procedure allows us to extract the $\gamma$-ray emissivities that are actually 
correlated with the \HI\ and CO components.

\subsection{IC and Point Sources}

To model $\gamma$-ray emission not related with interstellar gas, we referred to the GALPROP code 
\citep[e.g.,][]{Strong1998, Strong2007} for $\gamma$-rays produced through IC scattering
and to the first \textit{Fermi} LAT catalog (1FGL) for point sources \citep{Abdo2010_Catalog}.

GALPROP\footnote{\ \url{http://galprop.stanford.edu}} \citep{Strong1998,Strong2007} is a numerical
code which
solves the CR transport equation within the Galaxy and predicts the $\gamma$-ray emission produced 
via interactions of CRs with interstellar matter (nucleon-nucleon interaction and electron Bremsstrahlung) 
and low-energy photons (IC scattering). IC emission is calculated from the distribution of (propagated) 
electrons and the interstellar radiation fields developed by \citet{Porter2008}. 
Here we adopt the IC model map produced in the GALPROP run 54\_77Xvarh7S
in which the CR electron spectrum is adjusted to agree with that measured by the LAT
\citep{Abdo2009_CRE}. This GALPROP model has been used in publications
by the LAT collaboration such as \citet{Abdo2010_Pulsar}.

The 1FGL Catalog is based on the first 11 months of the science phase of the mission
and contains 1451 sources detected at a significance $\gtrsim 4\sigma$
(the threshold is 25 in term of test statistic, TS
\footnote{
\ The test statistic is defined as
$$
\mathrm{TS}=2(\ln L - \ln L_0)
$$
where $L$ and $L_0$ is the maximum likelihood with and without including the
source in the model, respectively. 
$L$ is conventionally calculated as
$\ln(L) = \Sigma_{i} n_{i} \ln(\theta_{i}) - \Sigma_{i} \theta_{i}$,
where $n_{i}$ and $\theta_{i}$ are the data and the model-predicted counts
in each pixel denoted by the subscript $i$, respectively
\citep[See, e.g.,][]{Mattox1996}.
TS is expected to be distributed as a
$\chi^2$ with $n-n_0$ degrees of freedom if the numbers of free parameters in
the model are respectively $n$ and $n_0$ (4 for sources in the 1FGL Catalog).
} 
).
For our analysis we considered 21 point sources in the ROI with TS larger than 50.

\subsection{Gamma-Ray Analysis Model}\label{anamodel}

Following a well-established approach that dates back
to the COS-B era \citep[e.g.,][]{Lebrun1983}, we modeled the $\gamma$-ray emission as a linear
combination of maps tracing the column density of the interstellar medium.
This approach is based on a simple, but very plausible
assumption: $\gamma$-rays are generated through interactions
of CRs and the interstellar gas, and the ISM itself is transparent to $\gamma$-rays.
Then, assuming that CR densities do not significantly vary over the
scale of the interstellar complexes under study and that CRs penetrate clouds uniformly to their cores
we can model the $\gamma$-ray intensities to first order as a linear combination of contributions
from CR interactions with the different gas phases in the various regions along each line of sight.

We also added the IC model map by GALPROP
and models for point sources taken from the 1FGL Catalog
as described in \S~2.2. To represent the extragalactic diffuse emission
and the residual instrumental background from misclassified CR interactions in the LAT detector, we also added an isotropic component.
CR interactions with ionized gas are not explicitly included in the model. 
The mass column densities of ionized gas are poorly known, but their contribution is generally lower ($\le 10\%$) 
than that of the neutral gas and its scale height is much larger 
($\sim 1$~kpc compared with $\sim 0.2$ kpc; \citealt{Cordes2002}). We therefore expect the
diffuse $\gamma$-ray emission originating from ionized gas
to be largely accommodated in the fit by other components with large angular scales, such as the isotropic and IC
ones, and to minimally impact the determination of the neutral gas emissivities.

Therefore, the $\gamma$-ray intensities $I_{\gamma}(l,b)$ (${\rm s^{-1}~cm^{-2}~sr^{-1}~MeV^{-1}}$) can be modeled as:
\par\nobreak\noindent
\begin{eqnarray}
\lefteqn{I_{\gamma}(l, b) =} \nonumber \\
& \sum^{4}_{i=1}q_{{\rm HI}, i} \cdot 
N({\rm H\,\scriptstyle{I}})(l, b)_{i}
+ \sum^{3}_{i=1}q_{{\rm CO}, i} \cdot
W_{\rm CO}(l, b)_{i} \nonumber \\
& 
+q_{\rm EBV} \cdot {\rm \mbox{E(B-V)}_{res}}(l, b)
+I_{\rm IC}(l, b) \nonumber \\
& +I_{\rm iso} + \sum_{j}{\rm PS}_{j}~~,
\end{eqnarray}
where sum
over $i$ represents the combination of the Galactic
regions,
$q_{{\rm HI},i}$ (${\rm s^{-1}~sr^{-1}~MeV^{-1}}$) and
$q_{{\rm CO} ,i}$ (${\rm s^{-1}~cm^{-2}~sr^{-1}~MeV^{-1}~(K~km~s^{-1})^{-1}}$) 
are the emissivities per \HI\ atom and per $W_{\rm CO}$ unit, respectively.
$q_{\rm EBV}$ (${\rm s^{-1}~cm^{-2}~sr^{-1}~MeV^{-1}~mag^{-1}}$)
is the emissivity per unit of the $E(B-V)_{\rm res}$ map (for which 
independent normalizations are allowed between the positive and
negative residuals; see \S~\ref{sec:procedure}).
$I_{\rm IC}$ and 
$I_{\rm iso}$ are the
IC model and isotropic background intensities (${\rm
s^{-1}~cm^{-2}~sr^{-1}~MeV^{-1}}$),
respectively, and ${\rm PS}_{j}$ represents
the point source contributions.
Compared to the EGRET study by \citet{Digel2001}, 
we use two additional maps to better
trace the ISM: the CO map in the Perseus arm and the 
$E(B-V)_{\rm res}$ map.

\section{Data Analysis}

\subsection{Observations and Data Selection}

The LAT on board the \textit{Fermi} Gamma-ray Space Telescope, launched on 2008 June 11,
is a pair-tracking telescope, detecting photons from $\sim 20$~MeV to more than 300~GeV.
Details on the LAT instrument and pre-launch expectations of the performance
can be found in \citet{Atwood2009}, and the on-orbit calibration is described in \citet{Abdo2009_Cal}.

Routine science operations with the LAT started on 2008 August 4.  
We have accumulated events from 2008 August 4 to 2010 February 4
to study diffuse $\gamma$-rays in our ROI.
During this time interval the LAT was operated in sky survey mode nearly all of the time,
obtaining complete sky coverage every two orbits and relatively uniform exposures over time.  
We used the standard LAT analysis software, the \emph{Science Tools} 
and selected events satisfying the standard low-background event selection 
\citep[the so-called \emph{Diffuse} class;][]{Atwood2009}.\footnote{
\ Data and software are publicly available from the Fermi Science Support Center
(\url{http://fermi.gsfc.nasa.gov/ssc/}). 
For this analysis we used the P6 \emph{Diffuse} selection
and the \emph{Science Tools} version \texttt{v9r16p0}.
}
We also required the reconstructed zenith
angles
of the arrival direction of photons to be less than $105\arcdeg$ and the center of the LAT field of view
to be within $52\arcdeg$ from the zenith, in order to reduce the contamination of photons from the Earth limb.
In addition, we excluded the period of time during which the LAT detected bright GRBs,
i.e., GRB080916C \citep{Abdo2009_GRB080916C}, GRB090510 \citep{Abdo2009_GRB090510},
GRB090902B \citep{Abdo2009_GRB090902B}, and GRB090926A \citep{Abdo2010_GRB090926A}.

\subsection{Analysis Procedure}
\label{sec:procedure}

The model described by Equation~(1) was fitted to the data using the \emph{Science Tools}, which take into account
the energy-dependent instrument point spread function and effective area. We have analyzed
the LAT data from 100~MeV to 25.6~GeV using 
13 logarithmically-spaced energy bands from 100~MeV to 9.05~GeV, and a single band above 9.05~GeV.
We then have compared the model and data
in each energy band
using a binned maximum-likelihood method with Poisson statistics (in $0.25\arcdeg \times 0.25\arcdeg$ bins);
we thus did not assume an a-priori spectral shape of each model component except for the IC emission.
For the other components the convolution with the instrument response functions 
was performed assuming an $E^{-2}$ spectrum, and the integrated intensities were allowed to vary in each narrow energy bin.
Changing the fixed spectral shape index over the range from $-1.5$ to $-3.0$ has a negligible effect on the 
obtained spectrum. 
In the highest energy band, we have set
both the normalization and the spectral index free to accommodate the wider bin width.
We used a post-launch response function, P6\_V3\_DIFFUSE, developed to account for the $\gamma$-ray
detection inefficiencies due to pile-up and accidental coincidence in the LAT \citep{Rando2009}.
We stopped at 25.6~GeV since the photon statistics does not allow
us to reliably separate different gas components above this energy.

We started with point sources detected with high significance (${\rm TS} \ge 100$)
in the 1FGL Catalog; we have 14 sources in our ROI for which the normalizations are set free.
We also included 8 sources lying just outside ($\le 5\arcdeg$) of the region boundaries, 
with all the spectral parameters fixed to those in the 1FGL Catalog.
As a starting point we used \HI\ maps prepared for $T_{\rm S}=125~{\rm K}$.
We added model components step by step as described below.

We first fitted the LAT data using Equation~(1)
without  the $E(B-V)_{\rm res}$ map and the CO map in the Perseus arm, 
and then included the CO map and confirmed that the fit improved significantly;
the test statistic summed over 14 bands with separate fits
in each band (i.e., 14 more free parameters)
is 187.6.
The $\gamma$-ray emission associated to the gas traced by CO 
in the Perseus arm is thus significantly detected by the LAT.

Next we included the $E(B-V)_{\rm res}$ map in the analysis.
We allowed the independent normalizations
between the positive part and the negative part of the $E(B-V)_{\rm res}$ map,
and found that the normalizations differ with each other.
We thus will use the independent normalizations hereafter.
We chose this model to better represent the LAT data and
constrain the CR distributions, and leave a detailed discussion
about the use of dust as ISM tracer to a dedicated paper.
The improvement of the fit is very significant: ${\rm TS}=1119.6$ for 28 more free parameters.
The correlation between the $E(B-V)_{\rm res}$ map and the $\gamma$-ray residual map obtained by the fit
without the $E(B-V)_{\rm res}$ map, shown in Figure~\ref{fig:EBV}, further supports the use of $E(B-V)_{\rm res}$ map
in our analysis.

\begin{figure}
\includegraphics[width=0.5\textwidth]{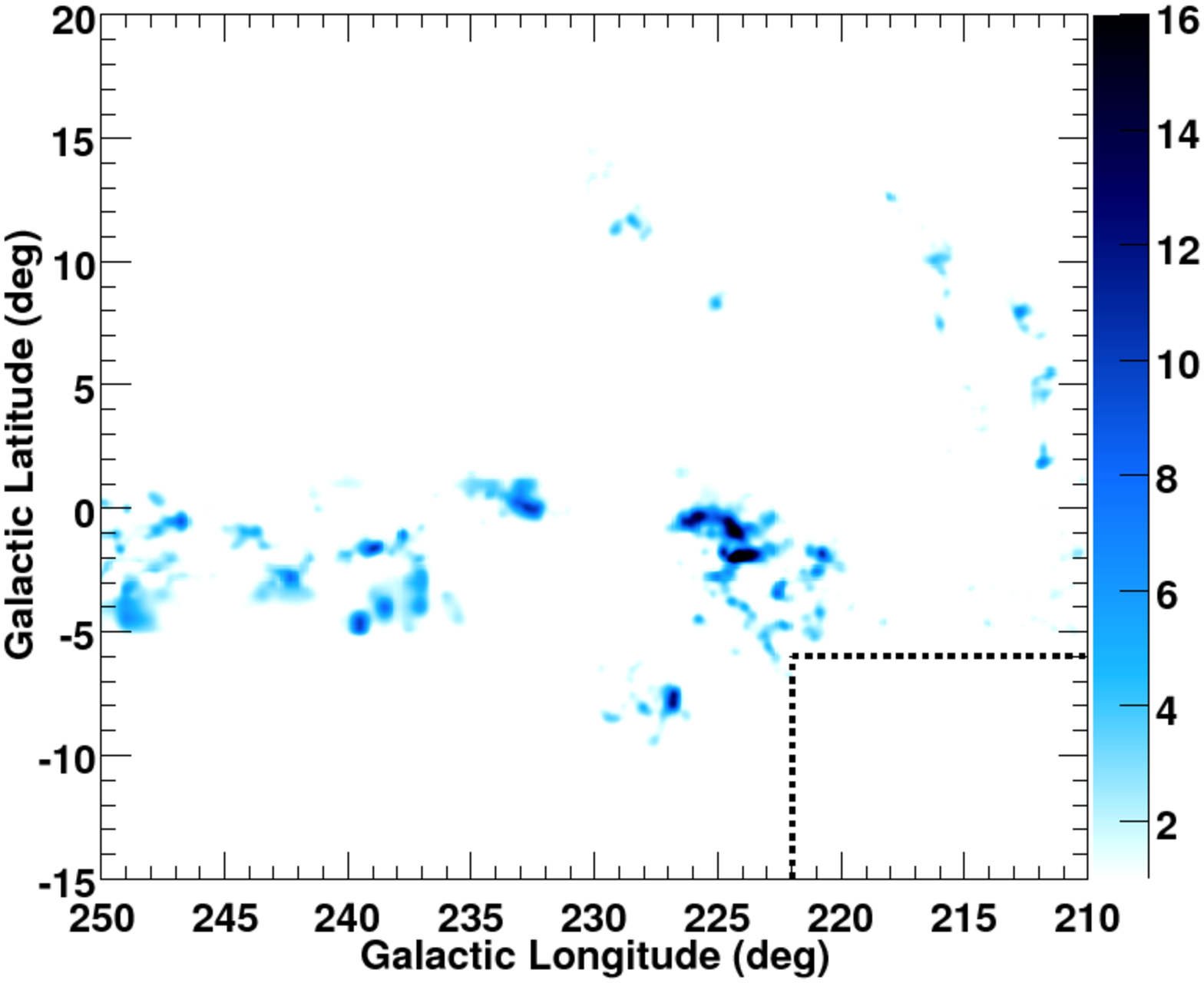}
\includegraphics[width=0.5\textwidth]{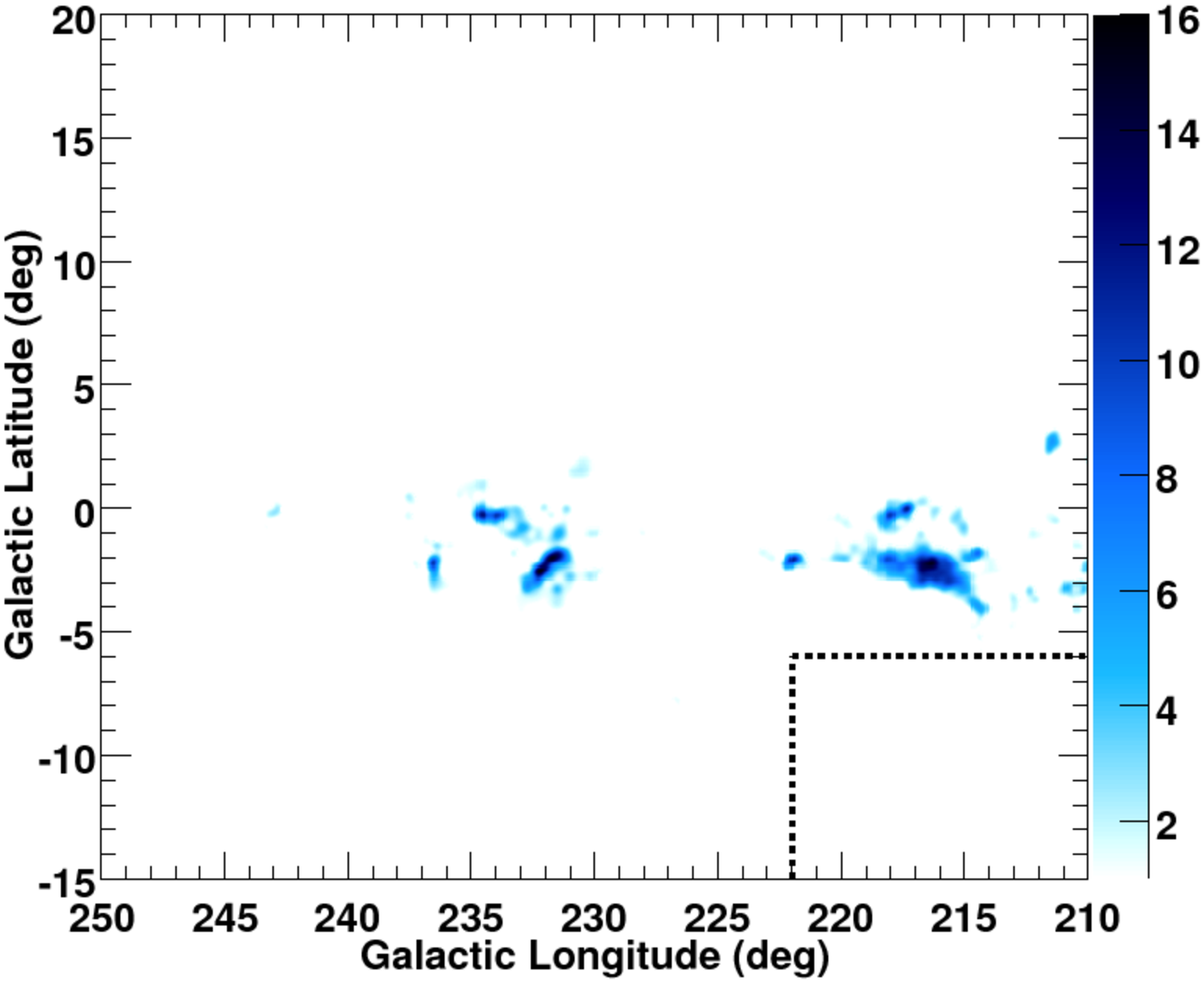}
\includegraphics[width=0.5\textwidth]{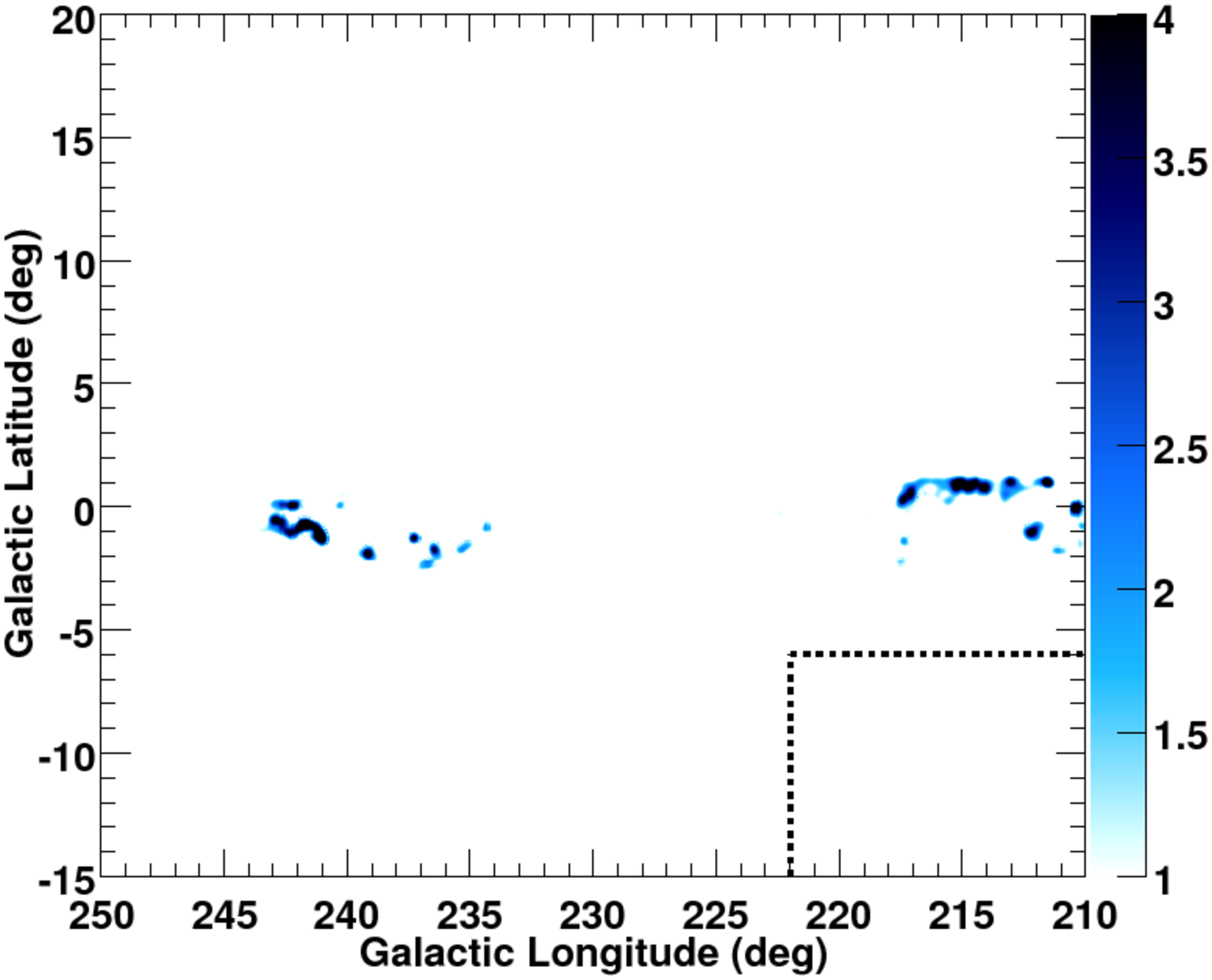}
\caption{
Maps of $W_{\rm CO}$ (in unit of ${\rm K~km~s^{-1}}$)
for the Local arm (top left), interarm (top right),
and the Perseus arm (bottom left) regions.
The small box in the bottom right corner indicates the area
not considered in the analysis.
The maps have been smoothed with a Gaussian of $\sigma = 0.25\arcdeg$ for display.
}
\label{fig:CO}
\end{figure}

\begin{figure}
\includegraphics[width=0.5\textwidth]{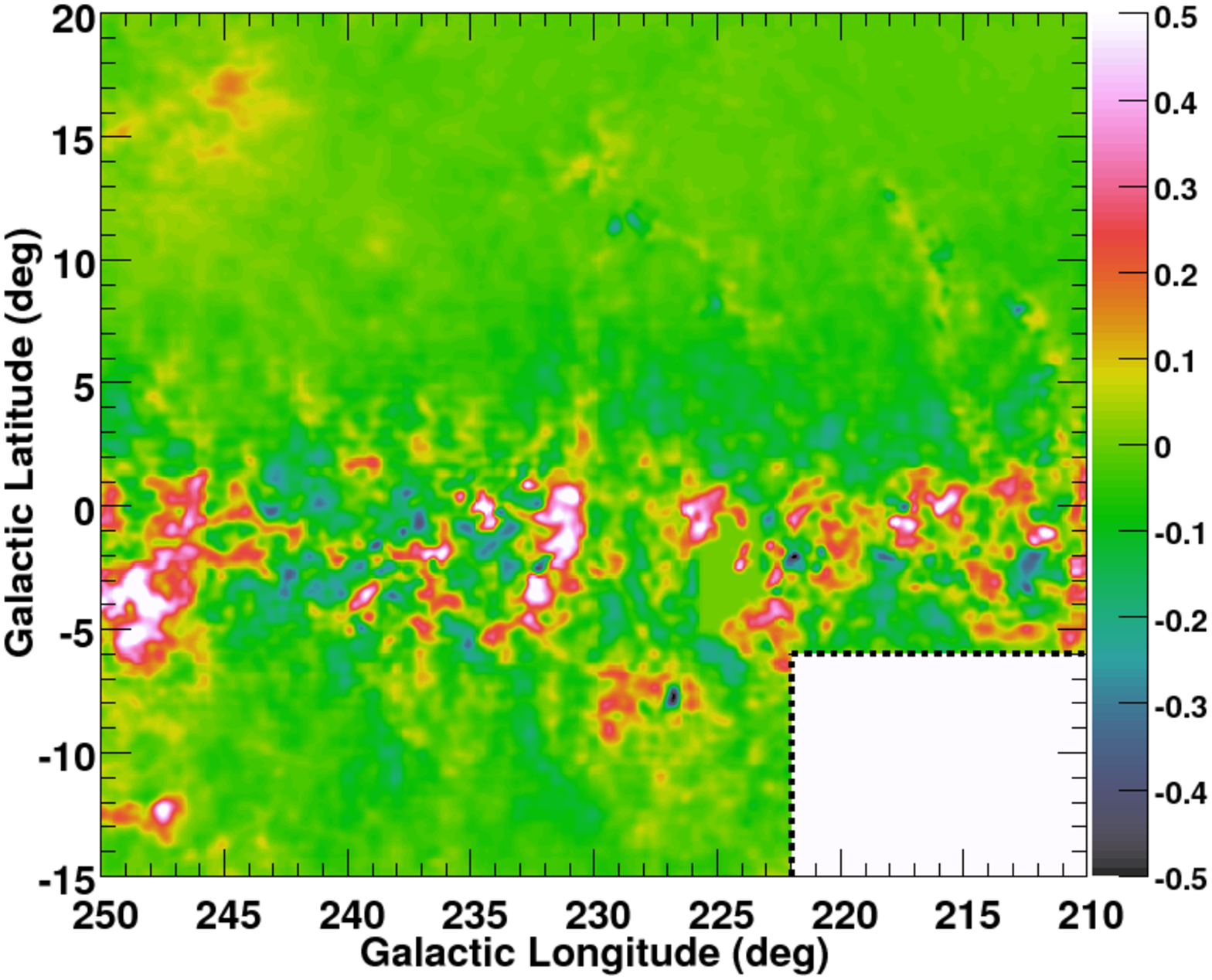}
\includegraphics[width=0.5\textwidth]{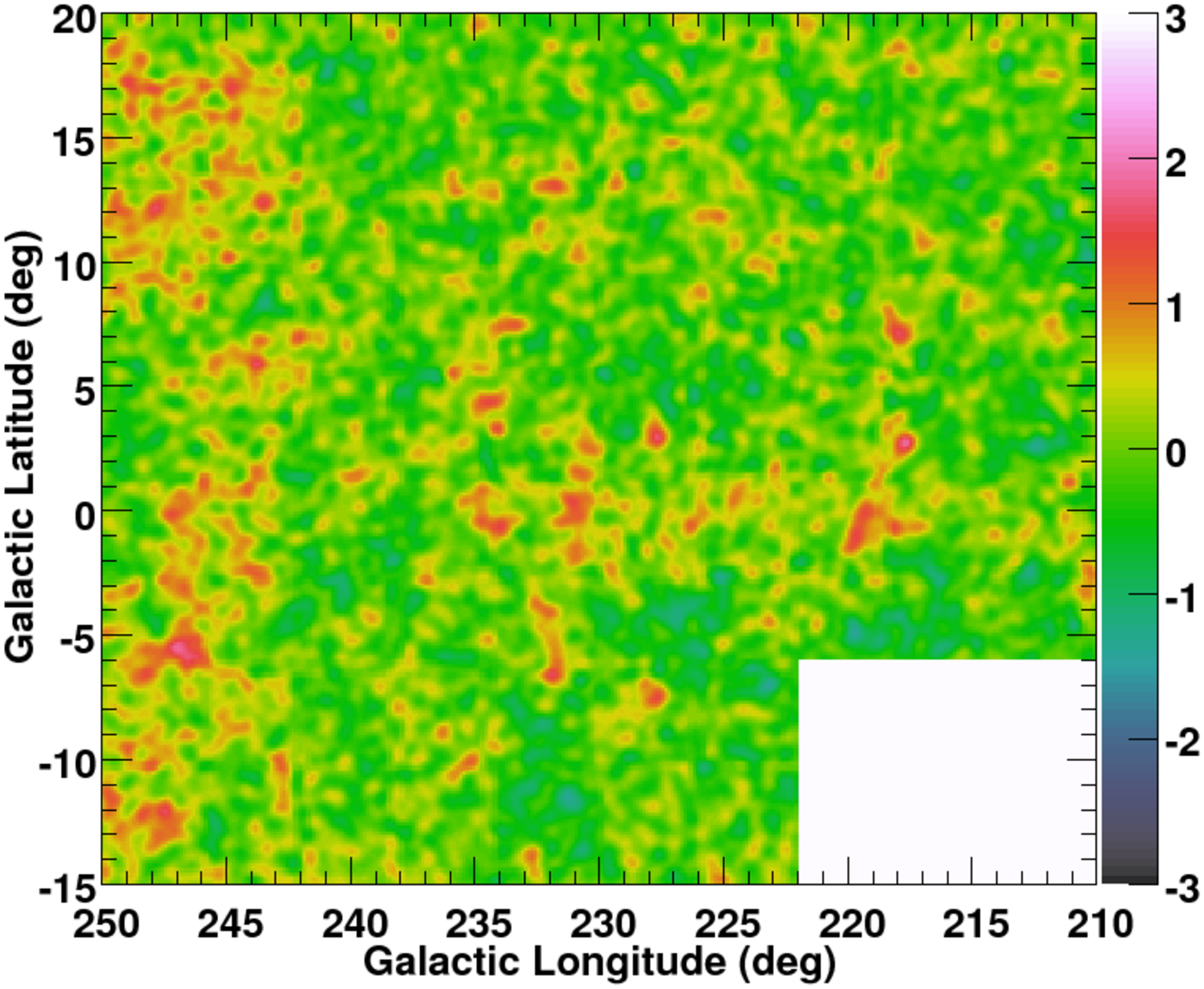}
\caption{
(left) Residual $E(B-V)$ map in unit of magnitudes,
obtained by subtracting the parts linearly correlated with the
combination of $N$(\HI) and $W_{\rm CO}$ maps.
The small box in the bottom right corner shows the area
not considered in the analysis.
The map has been smoothed with a Gaussian of $\sigma=0.25\arcdeg$ for display.
(right) $\gamma$-ray residual (data minus model) map obtained by the fit
without the $E(B-V)_{\rm res}$ map (only \HI\ and CO maps) in unit of standard deviations
(square root of model-predicted counts, saturated between $-3\sigma$ and $+3\sigma$).
The map has been smoothed with a Gaussian of $\sigma=0.5\arcdeg$.
}
\label{fig:EBV}
\end{figure}

We also tried a fit without the IC component to assess the systematics.
The effects on the derived emissivities are typically 2--3\% and $\sim$ 5\% for
$q_{\rm HI}$ and $q_{\rm CO}$, respectively.
They are much smaller than the statistical errors and
systematic uncertainties (see below), 
although the inclusion of the IC map improves the fit to the LAT data.
Therefore the uncertainties on the IC model have no significant 
impact on our analysis
due to its rather flat distribution across the region of interest
while the gas in the ISM is highly structured.
On the other hand, lowering the threshold for point sources down to 
${\rm TS=50}$ yields an about twice smaller emissivity
for the $W_{\rm CO}$ map in the Perseus arm. The
emissivities of other components are unchanged within the statistical errors.
This is plausibly due to the very clumpy distribution of the clouds in the Perseus arm as seen by a
terrestrial observer, see Figure~\ref{fig:CO}, which makes it 
difficult to separate from that of some discrete sources.
We thus use Equation~(1) with point sources detected at ${\rm TS \ge 50}$ in the 1FGL Catalog
\footnote{ 
\ Spectral parameters of point sources of TS=50-100 are fixed to those
given in the 1 FGL Catalog in the highest energy bin. (9.05--25.6~GeV)}
as our baseline model,
but we do not consider 
the highly uncertain CO emissivities in the Perseus arm for the discussion.

We summarize the results in Tables~\ref{tab:125K} and \ref{tab:250K} for 
$T_{\rm S}=125~{\rm K}$ and 250~K, respectively,
and the number of counts in each energy bin in Table~\ref{tab:counts}.
The differential emissivities are multiplied by $E^{2}$
where $E$ is the center of each energy bin in logarithmic scale. They are given for each model component.
We note that our isotropic term ($I_{\rm iso}$) includes the contribution of the instrumental
background and might partially account also for ionized gas (see \S~\ref{anamodel}),
thus it is significantly larger than the extragalactic diffuse emission reported by
\citet{Abdo2010_EGB}.

\begin{figure}
\includegraphics[width=0.5\textwidth]{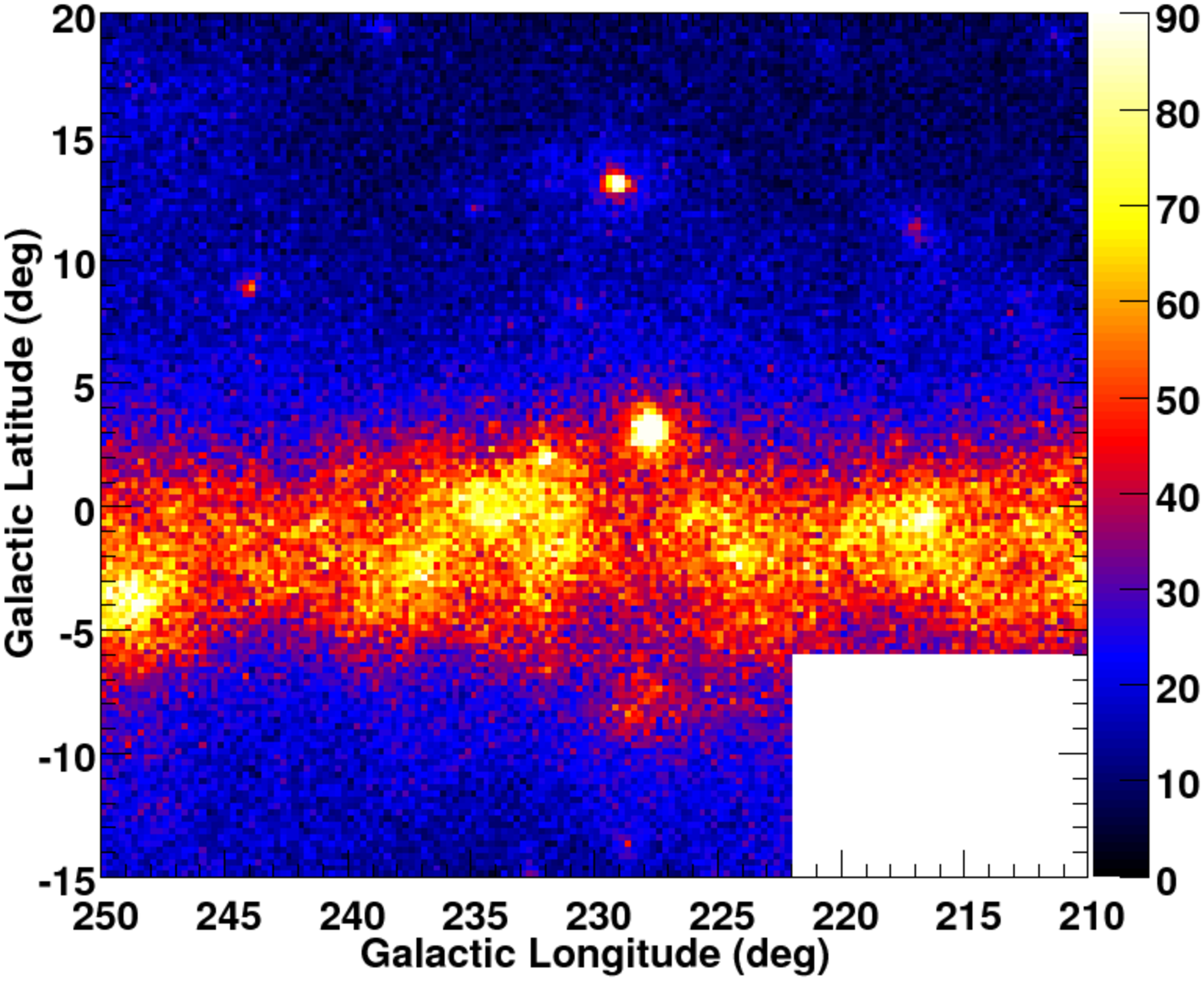}
\includegraphics[width=0.5\textwidth]{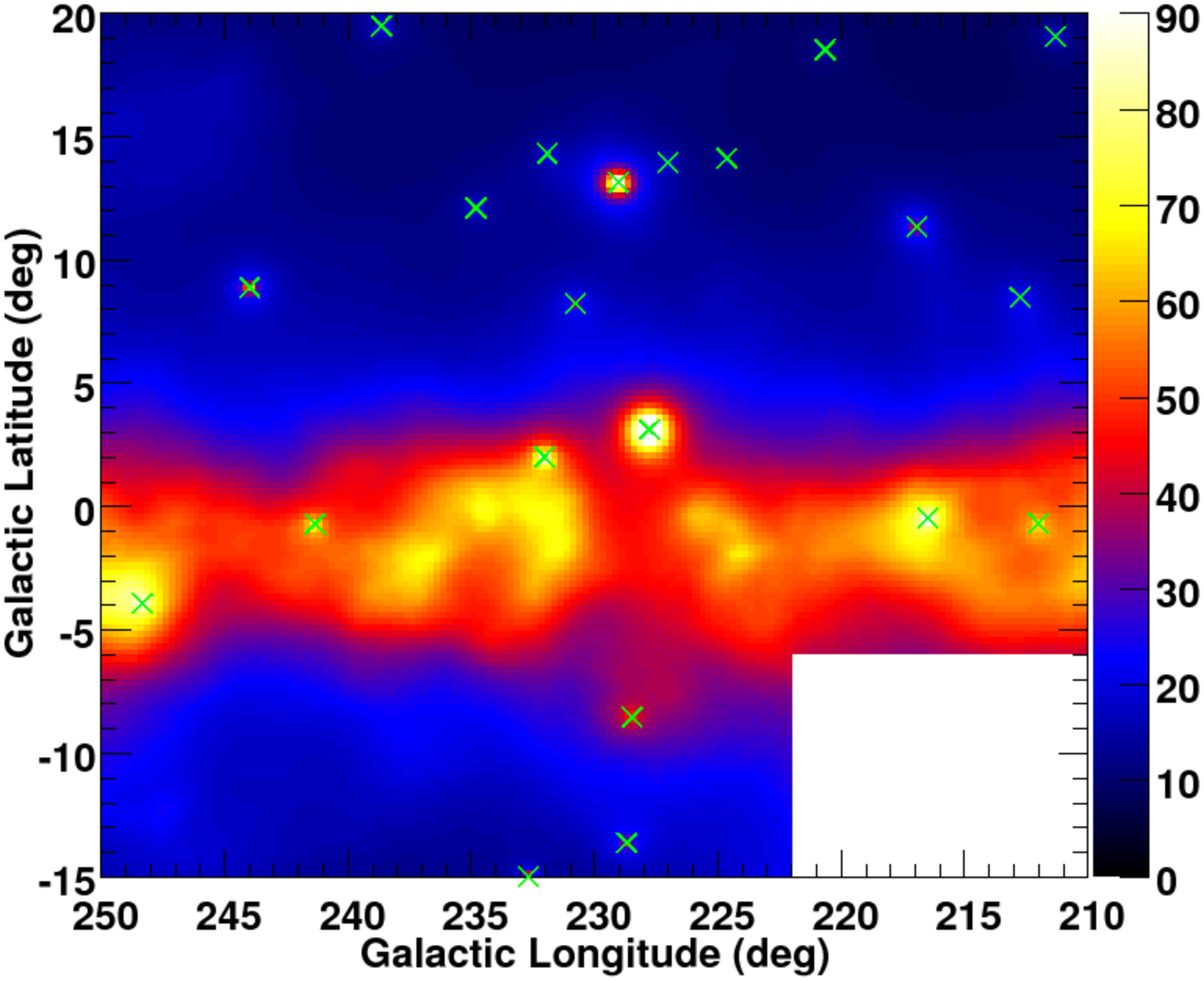}
\includegraphics[width=0.5\textwidth]{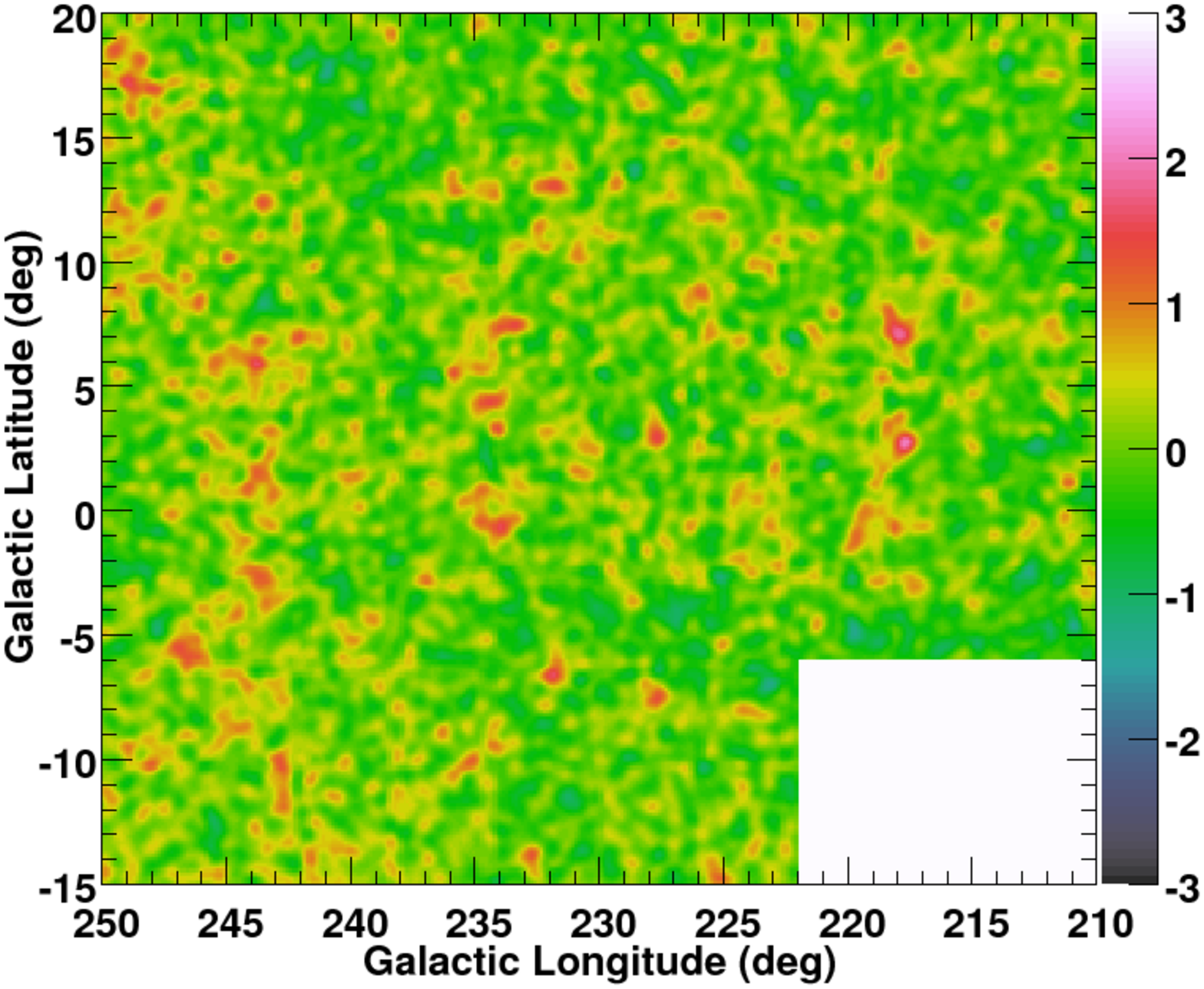}
\caption{
Data count map (top left), model count map (top right)
and the residual (data minus model) map in units of standard deviations (bottom left, saturated between $-3\sigma$ and
$+3\sigma$)
above 100~MeV obtained by our analysis. Point sources with
${\rm TS \ge 50}$ in the 1FGL included in the fit
are shown by crosses in the model map.
Data/model count maps are in $0.25\arcdeg \times 0.25\arcdeg$ pixels,
and the residual map has been smoothed with a Gaussian of $\sigma=0.5\arcdeg$.}
\label{fig:GammaMap}
\end{figure}

To illustrate the fit quality, we give the data and model count maps and the residual map 
in Figure~\ref{fig:GammaMap} (for $T_{\rm S}=125~{\rm K}$),
in which residuals (data minus model) are expressed in approximate standard deviation units 
(square root of model-predicted counts).
Although some structures (clustering of positive or negative residuals) are observed, 
the map shows no excesses below $-4~\sigma$ and above $6~\sigma$. Over 99\% of the pixels are within $\pm 3~\sigma$.
We thus conclude that our model reasonably reproduces the data.

\begin{figure}
\includegraphics[width=0.5\textwidth]{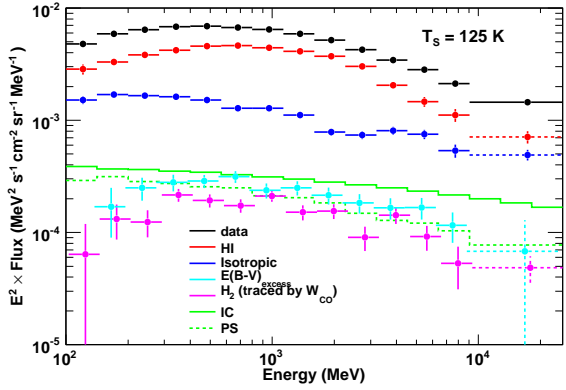}
\caption{
$\gamma$-ray spectra over the region of interest obtained from data and from the fitted model
(for each gas phase,
IC and isotropic components, and for point sources).
}
\label{fig:SummarySpectra}
\end{figure}

Figure~\ref{fig:SummarySpectra} presents the fitted spectra for each component.
The emission from the \HI\ gas dominates the $\gamma$-ray flux. Although the emission from the gas
in the CO-bright phase and that traced by $E(B-V)_{\rm res}$ is 
fainter than the IC and isotropic components, their characteristic spatial structures
(see Figure~\ref{fig:HI} and \ref{fig:CO}) 
allow their spectra to be reliably constrained.

\begin{figure}
\includegraphics[width=0.5\textwidth]{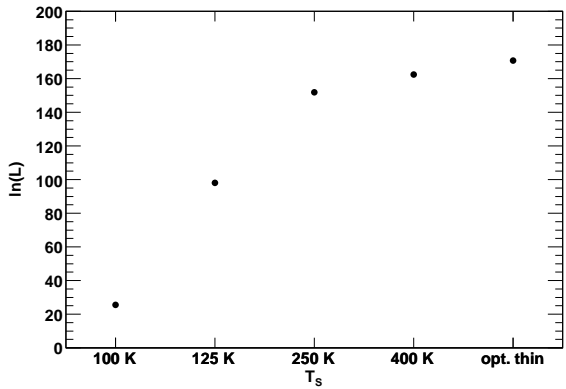}
\caption{
Variations of the log-likelihood value for several choices of $T_{\rm S}$ (the scale has a fixed
offset). 
}
\label{fig:logL}
\end{figure}

To examine the effect of the optical depth correction applied to derive the \HI\ maps,
as anticipated above we tried several choices of a uniform $T_{\rm S}$. 
We stress that the true $T_{\rm S}$ is likely to vary within clouds, but we stick to this simple
approximation exploring the following values:
100~K (which is a reasonable lower limit in the uniform approximation)\footnote{\
A truncation at 95~K was applied for channels where the brightness temperature was
larger.
},
250~K and 400~K (which are the two
values indicated by absorption measurements in the outer Galaxy by \citealt{Dickey2009})\footnote{\
Note that, however, the data used by
\citet{Dickey2009} do not cover the third Galactic quadrant}, 
and the optically-thin approximation 
(which yields the lower limit allowed on the atomic column densities).
The results on the maximum log-likelihood values
are summarized in Table~\ref{tab:lnL} together with the integrated \HI\ emissivities obtained above 100~MeV in
each region. 
The evolution of $\ln(L)$ with $T_{\rm S}$ is plotted in Figure~\ref{fig:logL}.
The \HI\ emissivity varies by $+15/\!-\!10 $\% for the Local arm, $+10/\!-\!0$\% for the interarm
region, 
and $+50/\!-\!25$\% for the Perseus arm with respect to the $T_{\rm S}=125~{\rm K}$ case.
We observe an increase of $\ln(L)$ with increasing spin temperature. 
Considering the fact that $T_{\rm S}=250~{\rm K}$ is a typical value 
in the second quadrant of the outer Galaxy according to a recent study by \citet{Dickey2009} and because
$\ln(L)$ saturates at $T_{\rm S} \ge 250~{\rm K}$, we regard 250~K as a plausible estimate of the average $T_{\rm S}$ in our ROI.
Unfortunately, the estimates by \citet{Dickey2009} have a rather large uncertainty 
(about~$ \pm 50$~K) in each Galactocentric radius bin, and they do not cover the region in
the third quadrant we are investigating \citep[see Figure~5 of][]{Dickey2009}.
In the following sections, we will concentrate on $T_{\rm S}=125~{\rm K}$ for comparison with previous
$\gamma$-ray measurements and on $T_{\rm S}=250~{\rm K}$ which agrees well with \HI\ absorption and the
LAT data.

\section{Results and Discussions}

\subsection{Emissivity Spectra of Atomic Gas}

\begin{figure}
\includegraphics[width=0.5\textwidth]{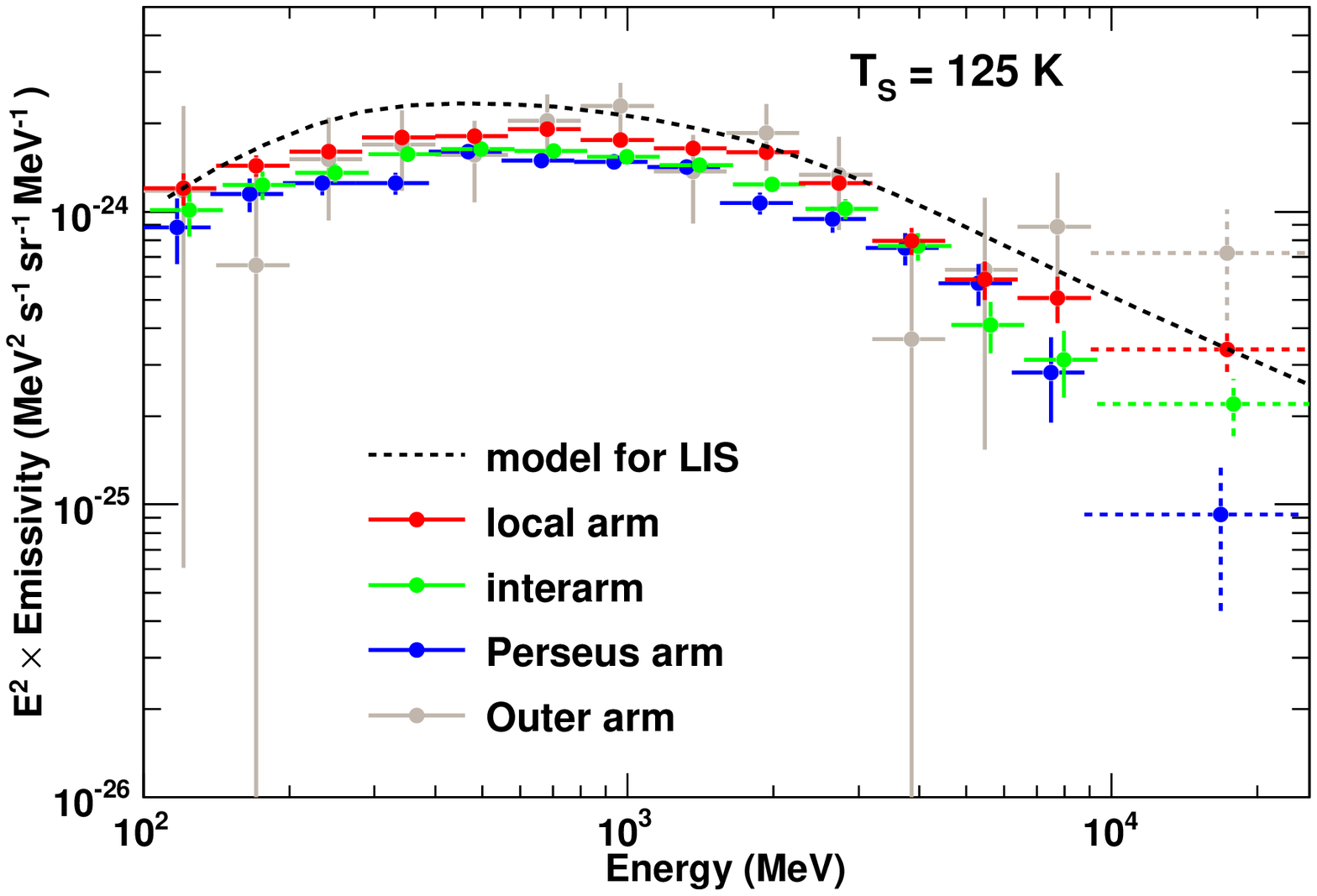}
\includegraphics[width=0.5\textwidth]{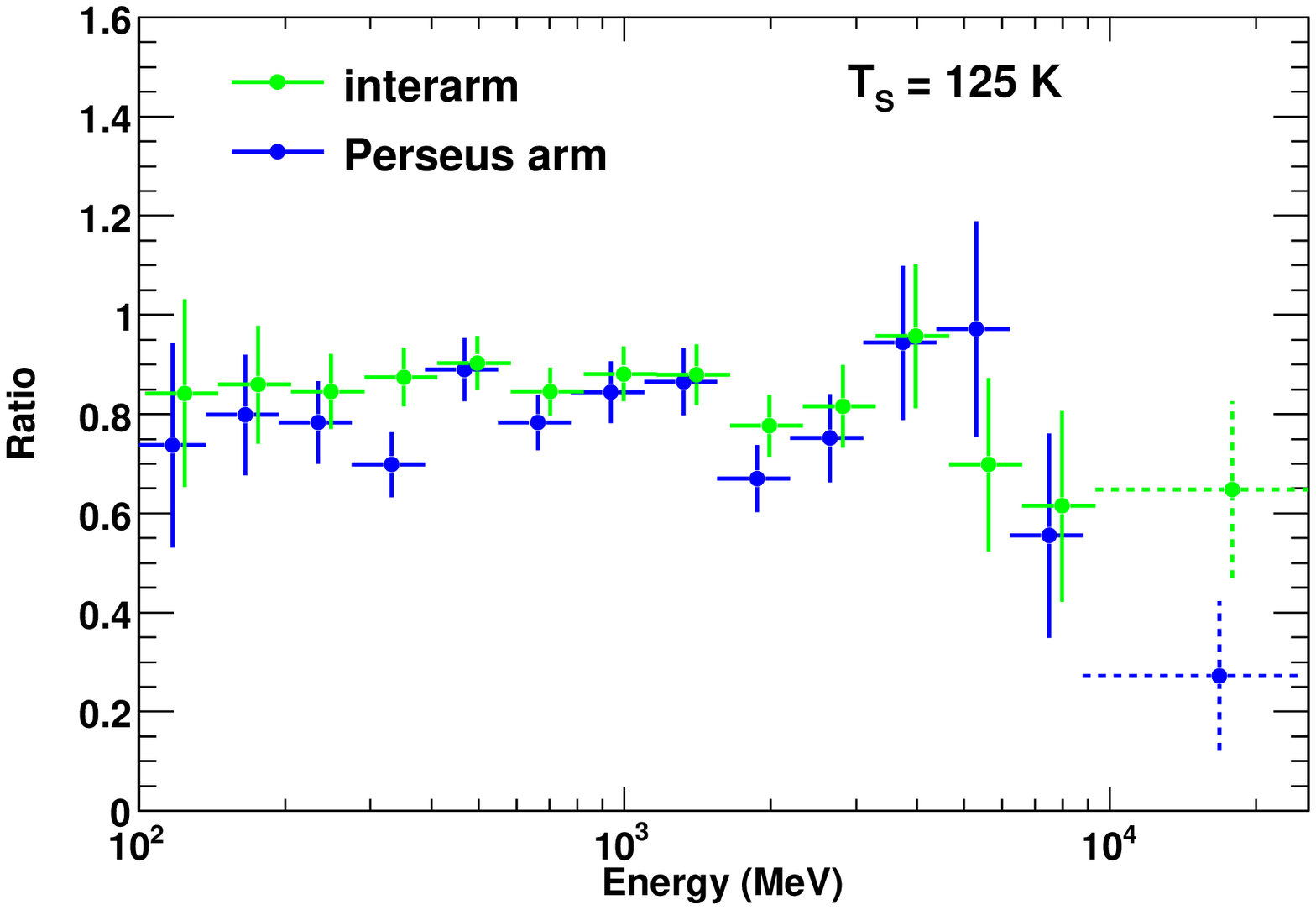}
\caption{
(left) \HI\ emissivity spectra obtained for each region. 
For reference,
the emissivity model spectrum for the LIS adopted by
\citet{Abdo2009_HI}
is shown by the black dotted line.
(right)
The emissivity ratios to that of the Local arm. 
In both panels, $T_{\rm S}=125~{\rm K}$ is assumed and
spectra in the interarm region and the Perseus arm
are shifted horizontally for clarity.
}
\label{fig:Spectra_125K}
\end{figure}

\begin{figure}
\includegraphics[width=0.5\textwidth]{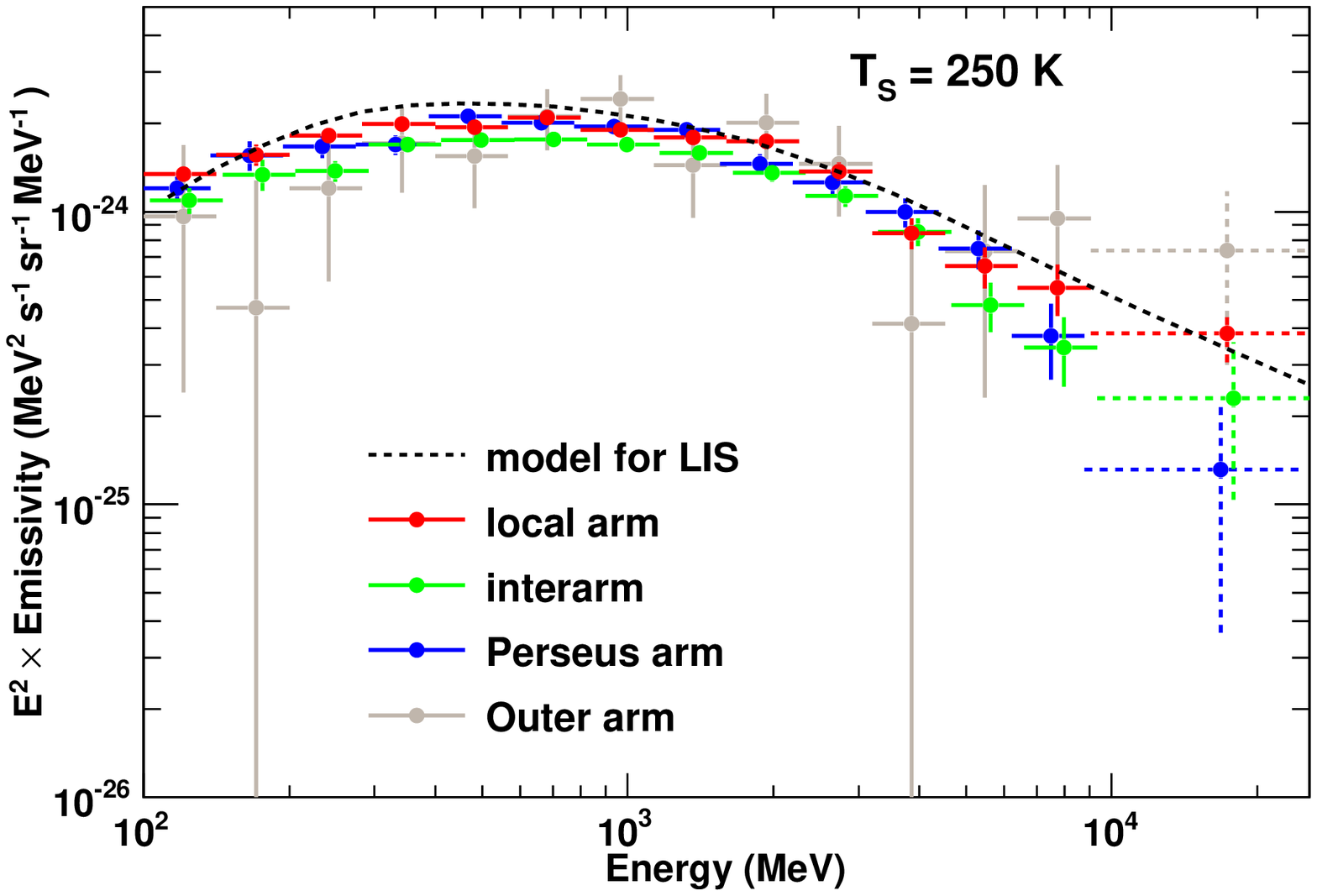}
\includegraphics[width=0.5\textwidth]{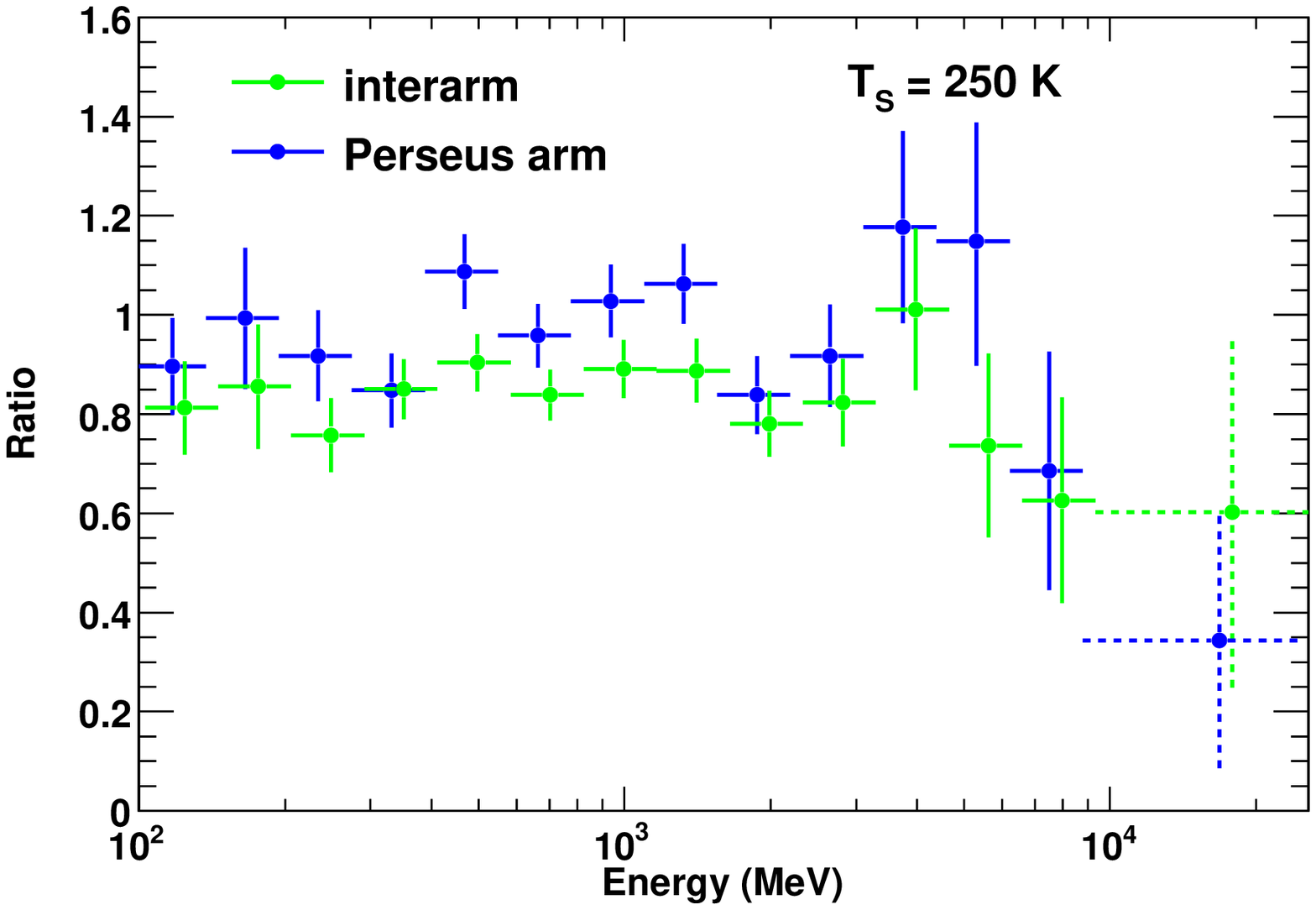}
\caption{
The same as Figure~\ref{fig:Spectra_125K} but for $T_{\rm S}=250~{\rm K}$.
}
\label{fig:Spectra_250K}
\end{figure}

In Figures~\ref{fig:Spectra_125K} and \ref{fig:Spectra_250K} (left panels), 
we report the emissivity spectra found per H atom in the Local arm, interarm, Perseus arm
and outermost regions 
for $T_{\rm S}=125$ and 250~K, respectively.
For comparison with the local interstellar spectrum (LIS) 
we also plot the model spectrum used in \citet{Abdo2009_HI} which agrees well 
with LAT data in a mid-to-high-latitude region ($22\arcdeg \le \left| b \right| \le 60\arcdeg$) 
of the third quadrant (assuming $T_{\rm S}=125~{\rm K}$). We see that 
the spectral shape of the Local-arm emissivity agrees well with the model for the LIS 
and does not depend on the choice of spin temperature. 
The integral emissivity of the Local arm is 10\% lower than that reported by \citet{Abdo2009_HI}
for the same spin temperature. This difference is not significant 
given the uncertainties
in the kinematic separation of the gas components. The present result is also consistent with the
measurement in the second quadrant \citep{Abdo2010_2nd}. Together they show that the CR density
along the Local arm is rather uniform within 1~kpc around the Sun, both in the second and third
quadrants.

The comparison of the data with the model emissivity expected for the Local arm region
based on locally-measured CRs (Figures~\ref{fig:Spectra_125K} and \ref{fig:Spectra_250K})
indicates a better fit for higher $T_{\rm S}$; 
$T_{\rm S}=125~{\rm K}$ gives emissivities 15--20\% lower than the model,
whereas $T_{\rm S}=250~{\rm K}$ shows better agreement by about 10\%.
Although the theoretical emissivity has uncertainties due to imperfect
knowledge of the CR spectrum \citep[see][]{Abdo2009_HI}, 
the fact that a high $T_{\rm S}$
value yields a better match both to the local absolute emissivity 
and to the spatial distribution of
the diffuse emission (Figure~\ref{fig:logL}) leads to larger $T_{\rm S}$ than 
a value conventionally used in $\gamma$-ray astrophysics (125~K).
This is in accord with independent estimates of $T_{\rm S}$ as discussed in
\S~\ref{sec:procedure}.

We also observe that the emissivity spectra do not vary significantly with Galactocentric distances in the outer Galaxy. 
To examine the spectral shape more quantitatively,
we present the emissivity ratios of the interarm and Perseus regions relative to the Local arm 
in the right panels of Figures~\ref{fig:Spectra_125K} and \ref{fig:Spectra_250K}.
The spectral shape in the interarm region is found to be consistent with that in the Local arm; 
a fit to the data for $T_{\rm S}=125~{\rm K}$ with a constant ratio
gives $\chi^2 = 7.3$ for 13 degrees of freedom. Although the fit is not fully acceptable for the
Perseus arm ($\chi^2 = 24.3$), 
the large $\chi^2$ is driven solely by the last bin. We note a possible interplay between the Perseus-arm
and the adjacent outer-arm emissivities in the highest energy bins (see left panels of Figure~\ref{fig:Spectra_125K} and \ref{fig:Spectra_250K}).
It can be due to a small but non-negligible spatial difference
between the modeled templates and data and/or to the presence of unresolved point sources 
(generally harder than diffuse emission). Photon fluctuations from the structured gas components 
can also lead the fit to a slightly different solution in the spatial separation of the components. 
One would expect these possible systematic uncertainties to become important at high energy 
given the limited counts in the overall map. It is difficult to quantitatively test these effects 
without knowledge of the true model distributions, but we can note that the small deviations seen at 
400--560~MeV and 1.6--2.2~GeV from a constant ratio are not confirmed
by the general trend of the other points. 
They indicate that there are systematic uncertainties not fully accounted
for by the statistical errors in the fit.
We thus do not claim nor deny the spectral softening of the Perseus arm
at high energy. 
A test using $T_{\rm S}=250~{\rm K}$ for the $N$(\HI) maps gives
the same conclusion on the spectral shape.
We thus conclude that the spectral shapes are consistent with the LIS in the $0.1-6$~GeV
energy band, independent of the assumed $T_{\rm S}$.
Considering that these $\gamma$-rays trace CR nuclei of energies from a few GeV 
to about 100~GeV \citep[see, e.g., Figure~\ref{fig:EmissivityGradient} of][]{Mori1997}, LAT data
indicate that the energy distribution of the main component of Galactic CRs 
does not vary significantly in the outer Galaxy in the third quadrant.
We note that \citet{Abdo2010_2nd} reported a possible spectral hardening in the Perseus arm in the second quadrant. 
This might be due to the presence
of the very active star-forming region of NGC~7538 and of CRs having not diffused far from their sources, or to contamination by
hard unresolved point sources. In fact \citet{Abdo2010_2nd} did not rule out the possibility 
that their result is due to systematic effects.

\subsection{Calibration of Molecular Masses}\label{xcopar}

High-energy $\gamma$-rays are a powerful probe to determine the CO-to-H$_2$ calibration ratio, $X_{\rm CO}$, 
if the CR flux is comparable in the different gas phases inside a cloud.
Since the $\gamma$-ray emission from the molecular gas is primarily due to CR interactions with ${\rm H_{2}}$, 
and since the molecular binding energy is negligible in processes leading to $\gamma$-ray production,
the emissivity per ${\rm H_{2}}$ molecule is expected to be twice the emissivity per \HI\ atom. 
Then, under the hypothesis that the same CR flux
penetrates the \HI-\ and CO-bright phases of an interstellar complex, we can calculate $X_{\rm CO}$ as
$q_{\rm CO} = 2 X_{\rm CO} \cdot q_{\rm HI}$.

\begin{figure}
\includegraphics[width=0.5\textwidth]{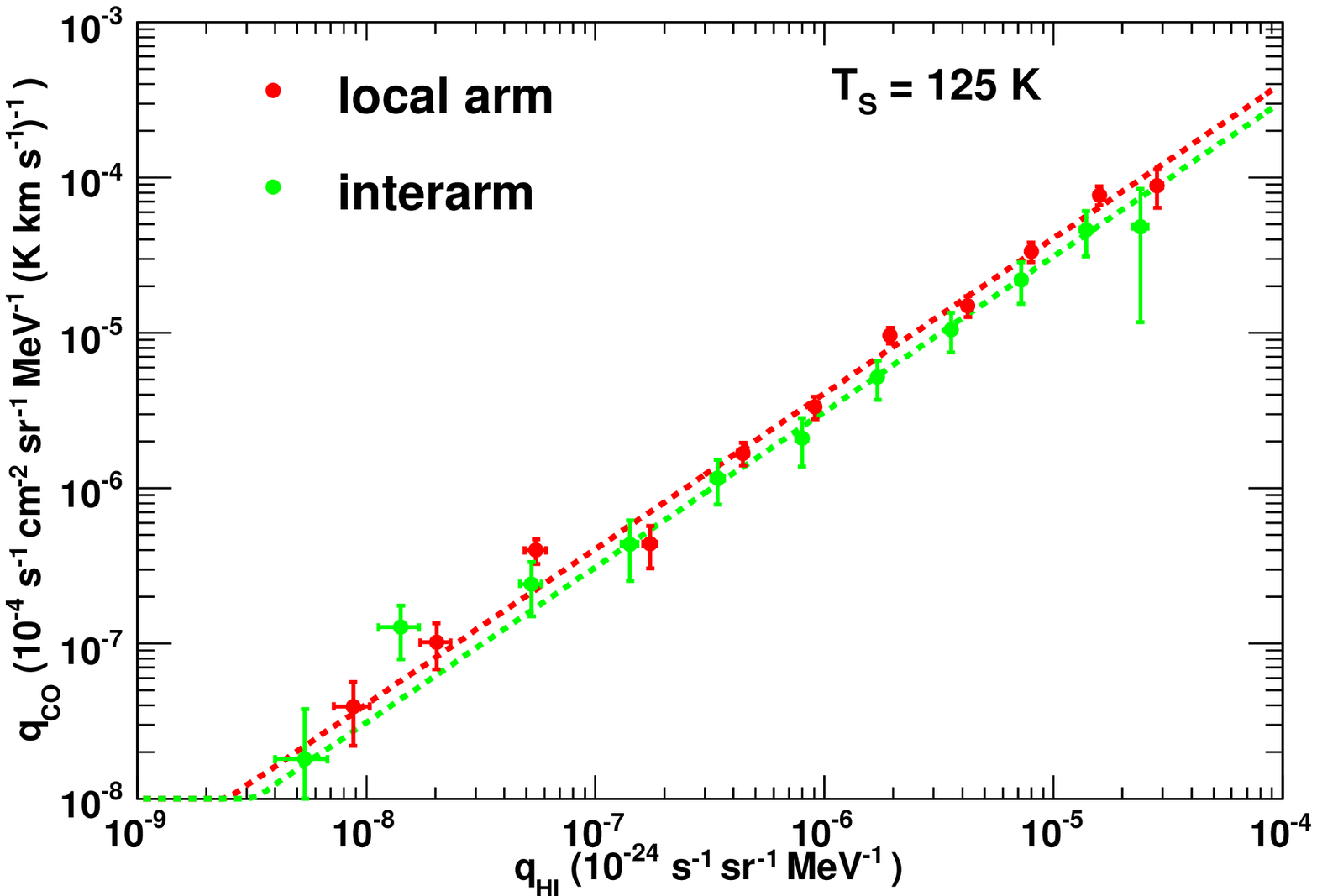}
\includegraphics[width=0.5\textwidth]{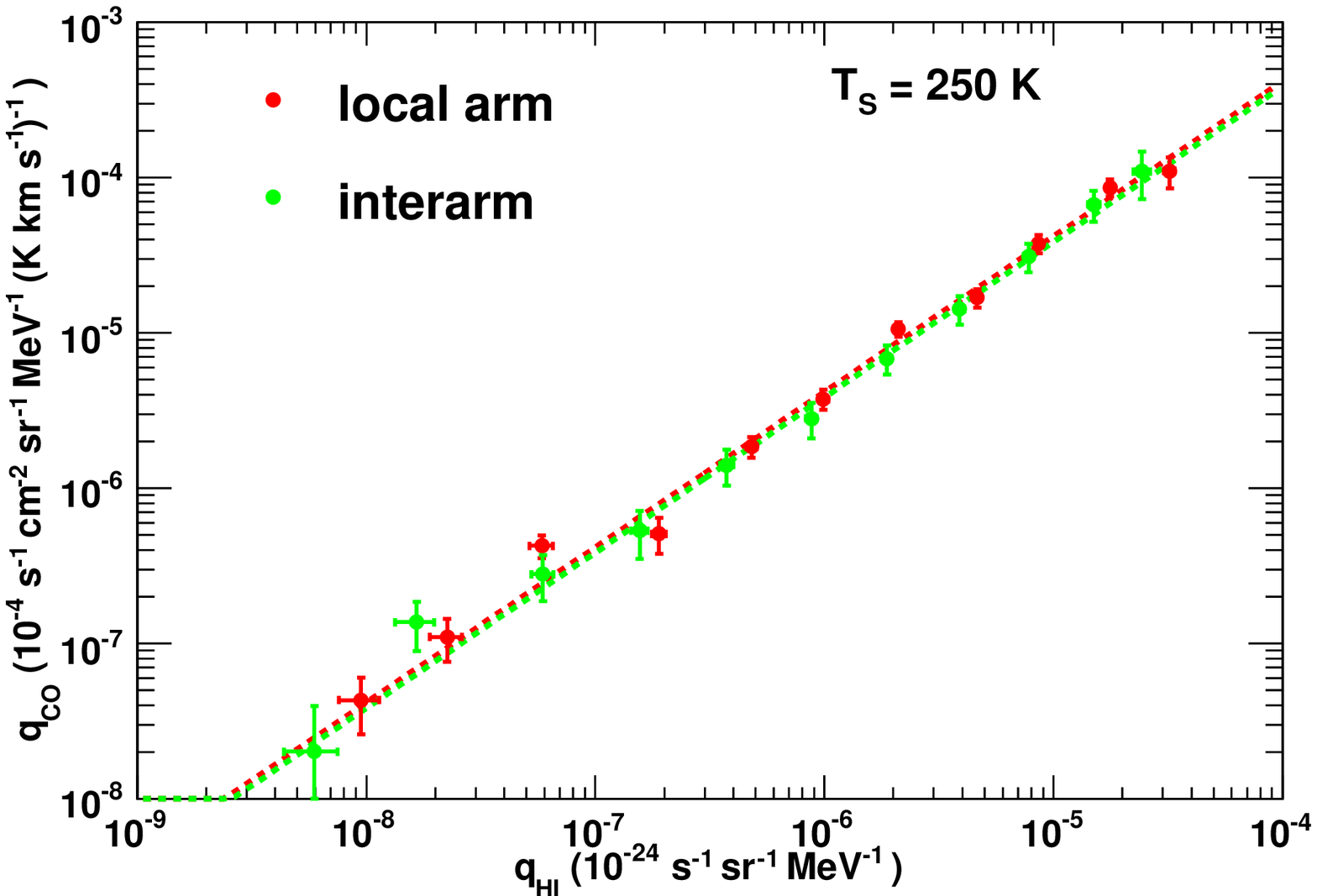}
\caption{
Correlation between the \HI\ and CO emissivities in the 200~MeV--9.05~GeV energy range 
for the Local arm and the interarm regions.
The cases of $T_{\rm S}=125~{\rm K}$ and 250~K
are shown in the left panel and the right panel, respectively.
Dotted lines show the best linear fits. 
Each data point corresponds to
an energy bin used in the $\gamma$-ray analysis. (See Table~\ref{tab:125K} and \ref{tab:250K})
}
\label{fig:Xco}
\end{figure}

We show $q_{\rm CO}$ as a function of $q_{\rm HI}$ for the
Local arm and the interarm region in Figure~\ref{fig:Xco}. 
We do not consider the correlation in the Perseus arm, because $q_{\rm CO}$ from this region
is affected by large systematic uncertainties (see \S~3.2).
Since the emissivity
associated with the CO-bright gas 
is not well determined in the lowest energy range (see Table~\ref{tab:125K} and \ref{tab:250K})
because of the poor angular resolution of the LAT,
and the fit at very high energy is affected by larger uncertainties (\S~4.1),
we have plotted only data in the 200~MeV--9.05~GeV range.
The linear correlation supports the assumption that Galactic CRs
penetrate molecular clouds uniformly to their cores.
It also suggests that contamination from point sources and CR spectral variations within the clouds are small.

We have derived the maximum-likelihood estimates of the slope and intercept of the linear relation 
between $q_{\rm CO}$ and $q_{\rm HI}$ taking into
account that $q_{\rm CO}$ and $q_{\rm HI}$ are both measured (not true) values with known uncertainties. 
The resulting intercepts are compatible with zero. The $X_{\rm CO}$ values we have obtained for $T_{\rm S}=250~{\rm K}$ are 
$(2.08 \pm 0.11) \times 10^{20}~{\rm cm^{-2}~(K~km~s^{-1})^{-1}}$ for the Local arm ($R \leq 10~{\rm
kpc}$) and
$(1.93 \pm 0.16) \times 10^{20}~{\rm cm^{-2}~(K~km/s)^{-1}}$ for the interarm region ($R=10\mbox{--}12.5~{\rm kpc}$).
Decreasing the spin temperature to 125~K does not affect the $X_{\rm CO}$ derivation in the well resolved, 
not too massive, clouds of the Local arm where we find 
$X_{\rm CO} = (2.03 \pm 0.11) \times 10^{20}~{\rm cm^{-2}~(K~km~s^{-1})^{-1}}$. On the other hand,
the separation in the $\gamma$-ray fit between the dense \HI\ peaks and clumpy CO cores becomes more
difficult for more distant, less resolved clouds where \HI\ and CO tend to peak in the same
directions. A change in the largest $N$(\HI) column-densities from the optical depth
correction can
impact the $X_{\rm CO}$ determination in two ways: first by changing the $q_{\rm HI}$ emissivity,
second by modifying the $N$(\HI) contrast within the cloud, hence the \HI\ and CO separation.
The global impact is mild since we find $X_{\rm CO} = (1.56 \pm 0.17) \times 10^{20}~{\rm
cm^{-2}~(K~km~s^{-1})^{-1}}$
in the interarm region for $T_{\rm S}=125~{\rm K}$.

\citet{Abdo2010_2nd} reported comparable values of $X_{\rm CO}$
in the second quadrant for $T_{\rm S}=125~{\rm K}$: they obtained
$(1.59 \pm 0.17) \times 10^{20}~{\rm cm^{-2}~(K~km~s^{-1})^{-1}}$ and
$(1.9 \pm 0.2) \times 10^{20}~{\rm cm^{-2}~(K~km/s)^{-1}}$ for the Local arm ($R \leq 10~{\rm kpc}$)
and the Perseus arm.
Given the systematic uncertainty in $X_{\rm CO}$,
roughly of the order of $0.3\times 10^{20}~{\rm cm^{-2}~(K~km~s^{-1})^{-1}}$, 
due to \HI\ optical depth correction, the results of both studies point to a rather uniform ratio over several kpc 
in the outer Galaxy.
Yet, these values are twice larger than found in the very nearby
Gould-Belt clouds of Cassiopeia and Cepheus, $X_{\rm CO} = (0.87 \pm 0.05) \times 10^{20}~{\rm
cm^{-2}~(K~km/s)^{-1}}$, $T_{\rm S}=125~{\rm K}$. However, we confirm that the
increase in $X_{\rm CO}$ beyond the solar circle is significantly lower than the trend adopted in the model of \citet{Strong2004b}. 
What fraction of the Gould-Belt to Local-arm difference in the average $X_{\rm CO}$ can be
attributed to a difference 
in the spatial sampling (resolution) of the clouds remains to be investigated.

Nearly the same region has been analyzed by \citet{Digel2001}
using EGRET data. The main difference from their analysis is our improved scheme for the kinematical separation 
of the ISM components along the lines of sight and the inclusion of the reddening residual map. 
The $X_{\rm CO}$ value measured in the Local arm by EGRET,
$(1.64 \pm 0.31) \times 10^{20}~{\rm cm^{-2}~(K~km~s^{-1})^{-1}}$ (for $T_{\rm S}=125~{\rm K}$),
is statistically compatible with ours.
The fact that we excluded the region of Monoceros R2 from the analysis
can also explain in part this difference.

Because of the pile-up of different clouds along a line of sight, the derivation of individual cloud masses 
is beyond the scope of this study. Let us just note that Maddalena's cloud, with its very low rate of star formation, 
seems to share a quite
conventional $X_{\rm CO}$ factor. Further investigation, including higher resolution $\gamma$-ray maps 
when more high-energy LAT data become available, 
is required to fully understand the mass distribution in the clouds.

\subsection{The Gradient of CR Densities beyond the Solar Circle}
\label{sec:CRgradient}

In Figure~\ref{fig:EmissivityGradient} (left panel) we show the emissivity
gradient found beyond the Solar circle for different spin temperatures. 
Here we do not include the results for the optically-thin approximation which is equivalent
to an infinitely high $T_{\rm S}$ and gives similar emissivities to
$T_{\rm S}=400~{\rm K}$.
The typical statistical
errors associated with these measurements are illustrated in the right panel for
the $T_{\rm S}=125~{\rm K}$ case. In the right panel, a shaded area shows the
characteristic systematic error due to the LAT event selection efficiency, evaluated to be
$\sim10\%$ in the energy range under study. 

In order to evaluate the impact of the delicate separation of the gas in the outermost region, 
we have compared two extreme cases. The first one adopts the kinematic $R = 16$ kpc boundary 
and applies no correction for velocity dispersion, the second assigns all the outer-arm gas to the Perseus arm.
The emissivity in the Perseus arm differs by about 5\% from the original one, and those
in the Local arm and interarm regions hardly change. Therefore these effects are significantly
smaller than the
uncertainties due to the optical depth correction of the \HI\ data.
We also note that the main effect of the LAT selection efficiency uncertainty is to rigidly shift the profile
without any significant impact on the gradient.

We thus conclude that the most important source of uncertainty in the CR density gradient derivation
is currently that in the $N$(\HI) determination.
This is mainly because the optical depth correction is larger for
dense \HI\ clouds in the Local and Perseus arms than for diffuse clouds
in the interarm region.
The loss in contrast between the dense
(low-latitude) and more diffuse (mid-latitude) \HI\ structures resulting from an increase in spin
temperature affects the fit, 
particularly in the Perseus component which is more narrowly concentrated near the plane. 
When probing the CR densities as the ``ratio'' between the observed numbers of
$\gamma$-rays to H atoms, 
at the precision provided by the LAT the uncertainties in the ISM densities are dominant.

\subsubsection{Comparison with EGRET and the arm/interarm
contrast}\label{contpar}

An interesting finding of the former EGRET analysis \citep{Digel2001} was
an enhancement of the $\gamma$-ray emissivity in the Perseus arm compared with the
interarm region. This possibility is relevant for models of
diffuse $\gamma$-ray emissions based on the assumption that
CR and ISM densities are coupled
\citep[e.g.,][and references therein]{Hunter1997}.

The Local arm emissivity obtained by the EGRET study for $T_{\rm S}=125~{\rm K}$ is $(1.81 \pm 0.17)
\times 10^{-26}~{\rm
photons~s^{-1}~sr^{-1}~H\mathchar`-atom^{-1}}$, which is $\sim 25\%$ larger than our LAT result.
However, the two studies are based on different \HI\ surveys which yield different total
$N$(\HI) 
column densities integrated along the lines of sight. The
column-density ratios between the surveys varies from 0.6 to 1.0 within the ROI, with an average value of
$\sim 0.8$. The difference is likely due to the improved correction for stray-radiation in the more recent survey, 
as discussed in \citet{Kalberla2005}.
The EGRET Local-arm emissivity scaled by 0.8 is in good agreement with our result for the same spin
temperature. 
If we do not include the $E(B-V)_{\rm res}$ map in the fitting, we obtain an emissivity of $(1.68 \pm 0.05) \times 10^{-26}~{\rm
photons~s^{-1}~sr^{-1}~H\mathchar`-atom^{-1}}$ which is still consistent with the down-scaled EGRET result within $\sim$15\%.
We thus conclude that our result is consistent with the previous study but is more reliable because of 
higher $\gamma$-ray statistics, finer resolution and an improved \HI\ gas survey.

\begin{figure}
\includegraphics[width=0.5\textwidth]{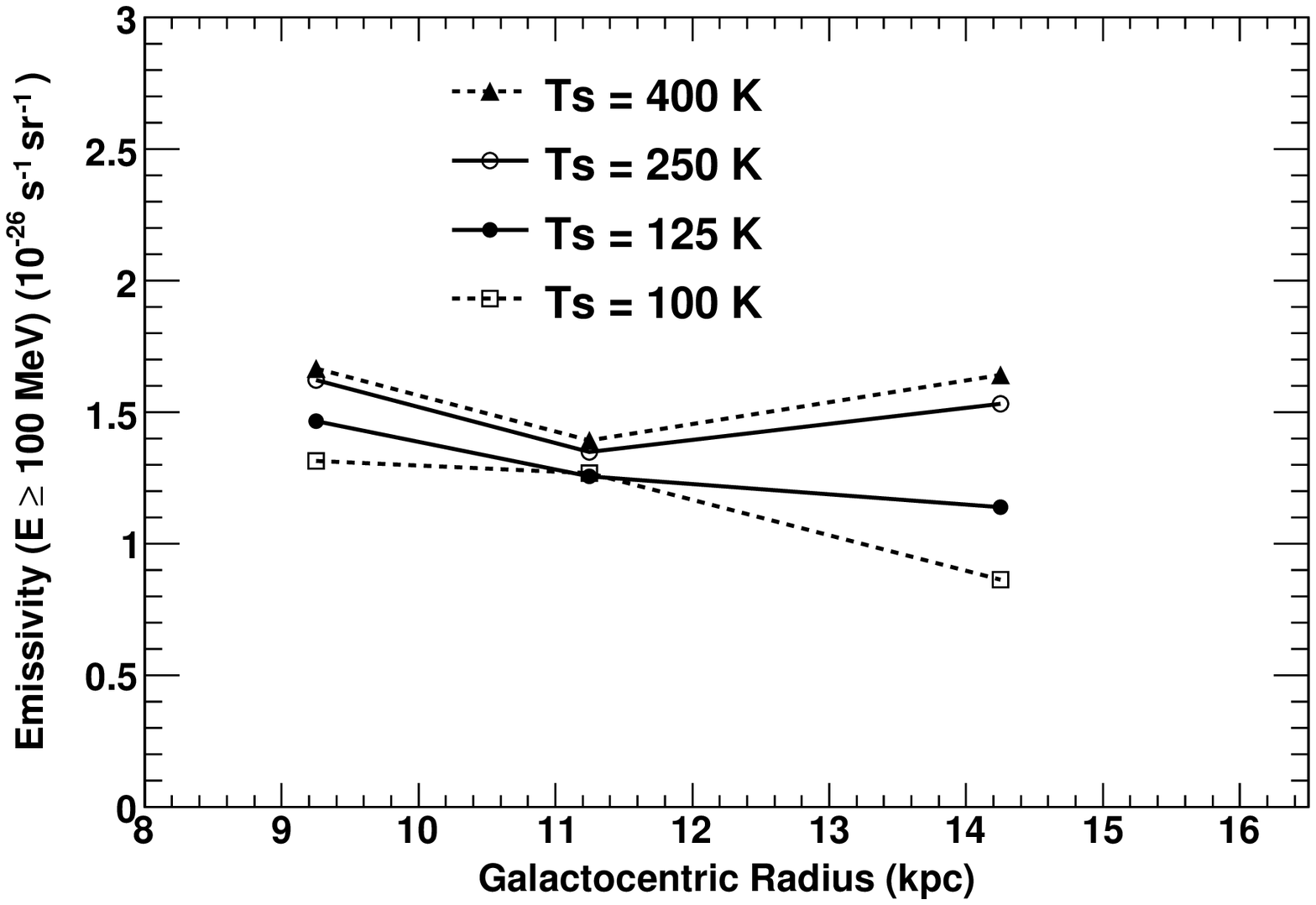}
\includegraphics[width=0.5\textwidth]{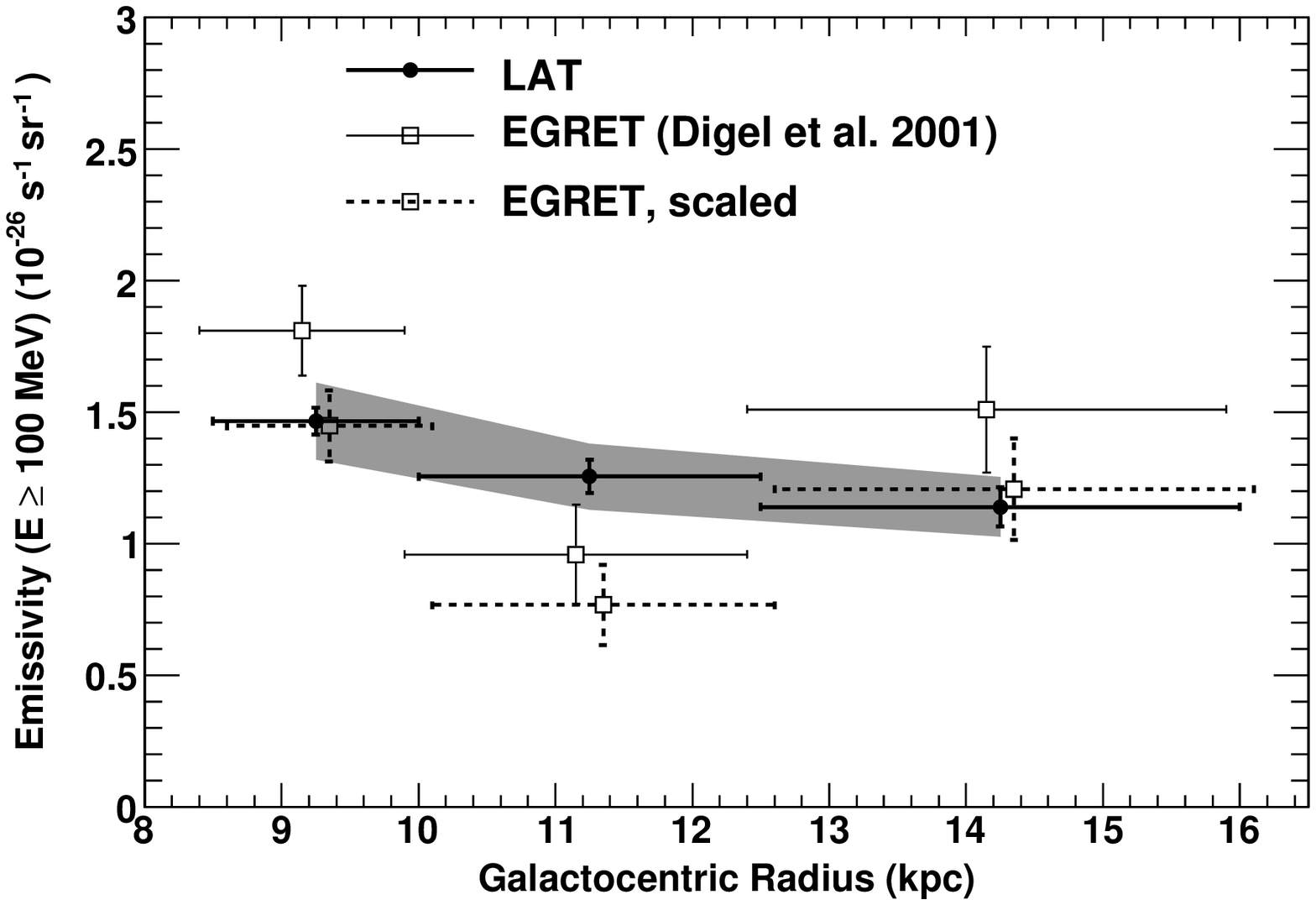}
\caption{
(left) Emissivity gradient for several choices of $T_{\rm S}$.
(right) Emissivity gradient obtained by the LAT
compared with the EGRET results under the assumption of $T_{\rm S}=125~{\rm K}$. 
The shaded area indicates the systematic
uncertainty in the LAT selection efficiency of $\sim 10$\%. The EGRET points have been downscaled
by 20\% to account for the change in \HI\ survey data between the two studies (see
\S~\ref{contpar}).
}
\label{fig:EmissivityGradient}
\end{figure}

We can therefore compare the present emissivity gradient (for consistency in
the case $T_{\rm S}=125~{\rm K}$) with that reported by the EGRET study, as summarized in
Figure~\ref{fig:EmissivityGradient} (right panel in which the EGRET results multiplied by 0.8 are also shown).
Although we observe good agreement between the two studies in the Local 
and the Perseus arms, this is not true for the arm/interarm contrast. The difference could be due to
the simple partitioning in cloud velocity used for the EGRET study. The \HI\ mass obtained for clouds in
the
interarm region with the simple partitioning is $20\%-40\%$ larger (for $T_{\rm S}=125~{\rm K}$) than with our
separation scheme, exaggerating the amount of gas in the
interarm region, and thus lowering the emissivity by the same amount.
Our emissivity profile is thus consistent with the previous study, but with improved precision (smaller statistical errors)
and accuracy (more reliable region separation method and better estimation of the point source contributions).
We thus do not confirm a marked drop in the interarm region.

Low spin temperatures yield a smooth decline in \HI\ emissivity to $R \simeq 16~{\rm kpc}$ in the outer Galaxy, without
showing a significant coupling with ISM column densities. The Perseus-to-interarm contrast is at
most of the order of 15\%--20\% for high spin temperatures
as shown in the left panel of Figure~\ref{fig:EmissivityGradient}. These profiles are similar at all
energies, in particular at high energies
where the component separation is more reliable thanks to the better angular resolution.
The surface density of \HI\ in the Perseus arm is on average 30\%--40\% higher
than in the interarm region. 
Therefore, even if we adopt $T_{\rm S}=400~{\rm K}$
which gives the largest arm-interarm contrast, the coupling scale
(or the coupling length) between the CRs and matter \citep[e.g.,][]{Hunter1997}
required to agree
with the LAT data would be larger than those usually assumed
for this type of model 
\citep[$\sim 2~{\rm kpc}$. See e.g.,][Figure~7]{Digel2001}.
Whether the true emissivity profile exhibits a small contrast between the arms or smoothly declines
with distance is beyond our measurement capability without further constraints on the \HI\
column-density derivation. New \HI\ absorption measurements will allow us to
investigate this issue with better accuracy.

\subsubsection{Comparison with a Propagation Model: the CR gradient problem}

To compare with the second quadrant results \citep{Abdo2010_2nd}, we have integrated the emissivities above 200~MeV 
for $T_{\rm S}=125~{\rm K}$. 
We find values of 
$(0.817 \pm 0.016) \times 10^{-26}~{\rm photons~s^{-1}~sr^{-1}~H\mathchar`-atom^{-1}}$, 
$(0.705 \pm 0.018) \times 10^{-26}~{\rm photons~s^{-1}~sr^{-1}~H\mathchar`-atom^{-1}}$ and
$(0.643 \pm 0.022) \times 10^{-26}~{\rm photons~s^{-1}~sr^{-1}~H\mathchar`-atom^{-1}}$ 
for the Local-arm, interarm, and Perseus-arm regions, respectively. 
The nearer value is about 20\% lower than in the 2nd quadrant 
(which, however, samples very nearby clouds in the Gould Belt) 
and the outer ones compare very well with the 2nd quadrant measurement 
over the same Galactocentric distance range. Despite the uncertainties due to the optical-depth correction 
(that might have a different impact in the two quadrants), 
both LAT studies consistently point to a slowly decreasing emissivity profile beyond $R = 10~{\rm kpc}$.

Let us consider the predictions by a CR propagation model to see the impact of such a flat profile 
on the CR source distribution and propagation parameters.
We adopted a GALPROP model, starting from the configuration used for the run
54\_77Xvarh7S which we used to predict the IC contribution. The CR source distribution in
this model is,
\par
\begin{figure}
\includegraphics[width=0.5\textwidth]{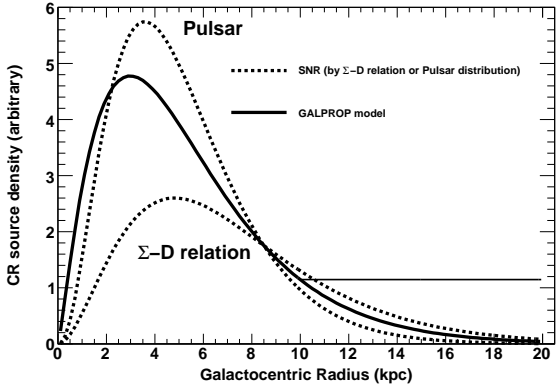}
\caption{
The CR source distribution adopted in our baseline GALPROP model (solid line), compared with the SNR
distribution
obtained by the $\Sigma-D$ relation \citep{Case1998}
and that traced by the pulsar distribution \citep{Lorimer2004}
shown by dotted lines.
The thin solid line represents an example of the modified distributions
introduced to reproduce the emissivity gradient
by the LAT.
}
\label{fig:CRsourceModel}
\end{figure}

\nobreak\noindent
\begin{equation}
f(R) = \left( \frac{R}{R_{\odot}} \right)^{1.25} \exp{\left(-3.56 \cdot \frac{R-R_{\odot}}{R_{\odot}}\right)}~~,
\end{equation}
where $R_{\odot}=8.5~{\rm kpc}$ is the distance of the Sun 
to the Galactic center. As shown in Figure~\ref{fig:CRsourceModel},
this function is intermediate between the distribution of supernova remnants (SNRs) 
obtained from the $\Sigma\mbox{--}D$ relation \citep{Case1998}
and that traced by pulsars \citep{Lorimer2004}.
The boundaries of the propagation region are chosen to be
$R_{\rm h}=30~{\rm kpc}$ (maximum Galactocentric radius) and 
$z_{\rm h}=4~{\rm kpc}$ (maximum height from the Galactic plane),
beyond which free escape is assumed. The spatial diffusion coefficient
is assumed to be uniform across the Galaxy and is taken as $D_{xx}=\beta D_{0} (\rho/{\rm 4~GV})^{\delta}$,
where $\beta \equiv v/c$ is the velocity of the particle relative to the speed of light and $\rho$ is the rigidity. 
We adopted $D_{0}=5.8 \times 10^{28}~{\rm cm^{2}~s^{-1}}$ and $\delta=0.33$ (Kolmogorov spectrum).
Reacceleration due to the interstellar magnetohydrodynamic turbulence,
which is thought to reproduce the observed B/C ratio at low energy, assumes an Alfv\'{e}n velocity $v_{\rm A}=30~{\rm km~s^{-1}}$.
The CR source distribution and propagation model parameters have been used often in the literature \citep[see e.g.,][]{Strong2004a}.
We note that the same CR source distribution and similar propagation parameters are
adopted in the GALPROP run used by \citet{Abdo2010_2nd}.

\begin{figure}
\includegraphics[width=0.5\textwidth]{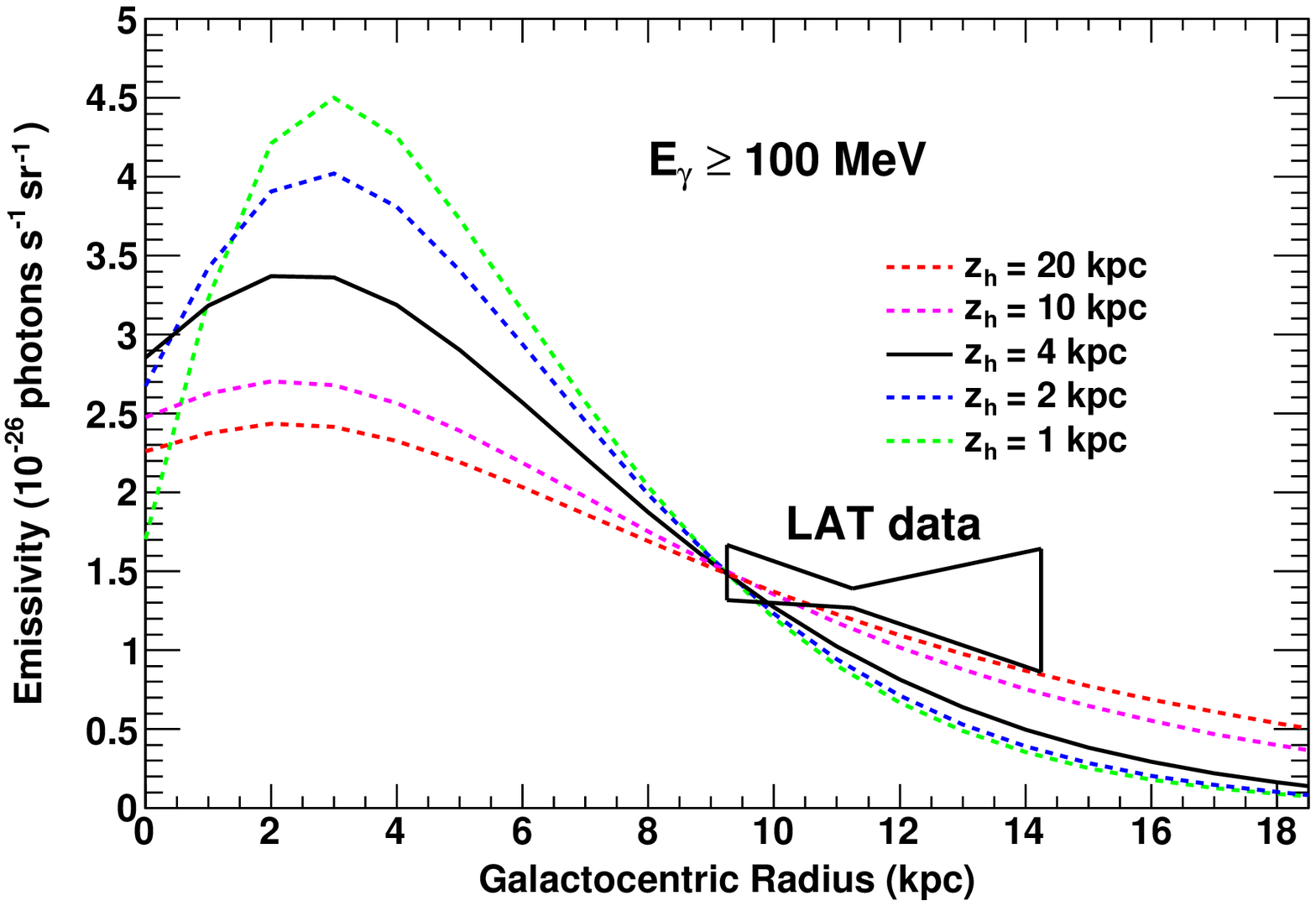}
\includegraphics[width=0.5\textwidth]{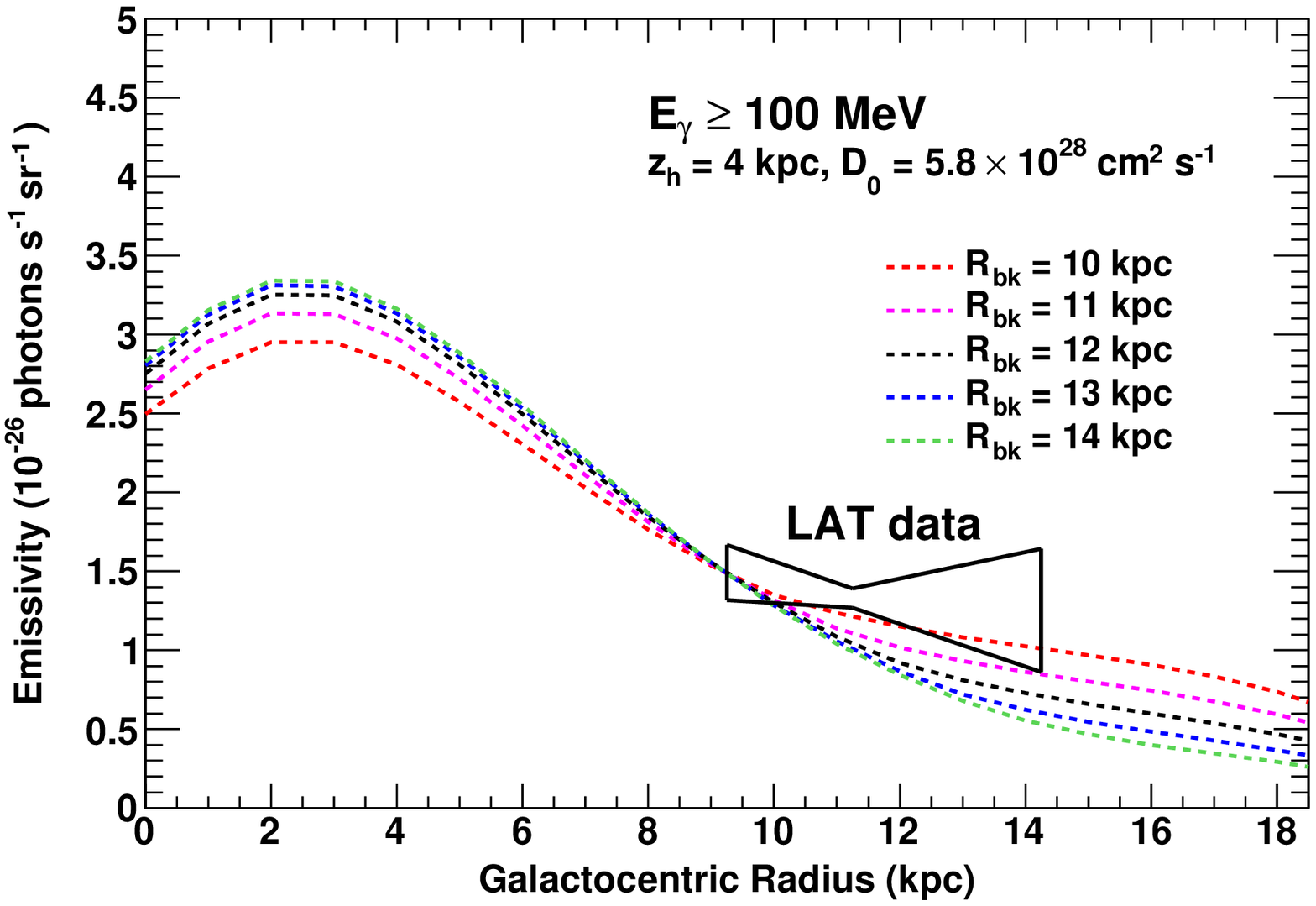}
\caption{
Comparison of the emissivity gradient obtained by the LAT and model expectations using GALPROP.
The left panel shows models with different halo sizes and diffusion lengths: ($z_{\rm h}, D_{0}$) =
(1~kpc, $1.7 \times 10^{28}~{\rm cm^{2}~s^{-1}}$),
(2~kpc, $3.2 \times 10^{28}~{\rm cm^{2}~s^{-1}}$),
(4~kpc, $5.8 \times 10^{28}~{\rm cm^{2}~s^{-1}}$),
(10~kpc, $12 \times 10^{28}~{\rm cm^{2}~s^{-1}}$) and
(20~kpc, $18 \times 10^{28}~{\rm cm^{2}~s^{-1}}$).
The solid line is for $z_{\rm h}=4~{\rm kpc}$.
The right panel shows different choices of the break distance
beyond which a flat CR source distribution is assumed:
$R_{\rm bk}=10\mbox{--}14~{\rm kpc}$ with 1~kpc steps.
}
\label{fig:GradientModelData}
\end{figure}

The left panel of Figure~\ref{fig:GradientModelData} compares the calculated profile
(solid line) with LAT constraints (bow-tie plot bracketing the profiles obtained for different $T_{\rm S}$;
see the left panel of Figure~\ref{fig:EmissivityGradient}).
The model is normalized to the LAT measurement in the innermost region. Despite the large uncertainties, 
LAT data lead to a significantly flatter profile than predicted by our model; the LAT 
results indicate to 
a factor of 2 larger emissivity (CR energy density) in the Perseus arm even if we assume $T_{\rm S}=100~{\rm K}$.
The higher $T_{\rm S}$ makes the discrepancy larger, hence the conclusion is robust.
Not using the $E(B-V)_{\rm res}$ map in the analysis does not change the conclusion, 
since the emissivities in the interarm region and the Perseus arm are almost unaffected by its presence.

The discrepancy between the $\gamma$-ray emissivity gradient in the Galaxy and the
distribution of putative CR sources has been known as the ``gradient problem'' since the COS-B era
\citep[e.g.,][]{Bloemen1989}. It has led to a number of possible interpretations,
including, for the specific case of the outer Galaxy, the possibility of a very steep gradient in
$X_{\rm CO}$ beyond the solar circle \citep[][]{Strong2004b}.
The emissivities in the outer Galaxy were more difficult to determine
in the COS-B/EGRET era due to lower statistics and higher backgrounds. Now, thanks to the
high
quality of the LAT data and the improved component separation technique applied to gas line
data, we measure
a flat \HI\ emissivity gradient in the outer Galaxy together with a flat evolution of $X_{\rm CO}$ 
over several kpc, so the gradient problem requires another explanation.

The most straightforward possibility is a larger halo size ($z_{\rm h}$), as discussed by,
e.g., \citet{Stecker1977}, 
\citet{Bloemen1989} and \citet{Strong1998}. We therefore tried several choices of
$z_{\rm h}$ and $D_{0}$ as summarized in the dotted lines in the same figure.
The values of $D_{0}$ are chosen to reasonably reproduce the LIS of protons and electrons,
B/C ratio and $\rm {^{10}Be/^{9}Be}$ ratio
at the solar system, and are similar to those given in \citet{Strong1998}.
All models are normalized to the LAT data in the Local arm. 
Models with $z_{\rm h} = 4~{\rm kpc}$ or smaller are found to give
too steep emissivity gradients. A CR source distribution as in equation~2 with
a very large halo ($z_{\rm h} \geq 10~{\rm kpc}$) provides a gradient compatible with the $\gamma$-ray data,
if we fully take into account the systematic uncertainties.
We note that $z_{\rm h} = 10~{\rm kpc}$ is still compatible with $\rm {^{10}Be/^{9}Be}$ measurements
\citep[e.g.,][]{Strong1998}.

Considering the large statistical and systematic uncertainties in the SNR distribution,
a flatter CR source distribution in the outer Galaxy also could be possible. 
We thus tried a modified CR source distribution, in which the distribution is the same as equation~2 below $R_{\rm bk}$ and
constant beyond it (see a thin solid line of Figure~\ref{fig:CRsourceModel} as an example).
Figure~\ref{fig:GradientModelData} right shows the models with several choices of $R_{\rm bk}$ for $z_{\rm h}=4~{\rm kpc}$
and $D_{0}=5.8 \times 10^{28}~{\rm cm^{2}~s^{-1}}$. We obtained a reasonable fit to the data using 
a flat CR source distribution beyond $R=10~{\rm kpc}$. Such a constant CR source density in the outer Galaxy is in
contrast not only with the (highly uncertain) distribution of SNRs, but
also with other tracers of massive star formation and SNRs, like,
1) CO lines which trace the interstellar phase where massive stars form \citep[e.g.,][]{Ferriere2001}, 
2) OB star counts \citep[e.g.,][]{Bronfman2000}, and
3) the $^{26}$Al line which is related to the injection
of stellar nucleosynthesis products in the ISM by SNRs
\citep{Diehl2006}.
However, a very large halo size and/or a flat CR source
distribution just beyond the solar circle
seem to be favored by the LAT data.

The above discussion depends on the propagation parameters and the solution is not unique. 
The exploration could be extended to other regions of
the parameter space or to a non-uniform diffusion coefficient \citep[e.g.,][]{Evoli2008}, but
examining propagation models in detail is beyond the scope of our study.
Our bottom line is that the analysis of LAT data presented here and by \citet{Abdo2010_2nd}
consistently show that the CR density gradient in the outer Galaxy
is flatter than expectations by commonly-used propagation models.
In the future, the extension to the inner part and the accurate determination of the gradient
over the whole Galaxy will be key to constraining the CR origin and transport.

We also note that a spin temperature $T_{\rm S} \ge 250~{\rm K}$, 
which is favored by
recent studies in the outer Galaxy \citep[e.g.,][]{Dickey2009}, gives a small arm/interarm contrast 
at the 10--20\% level that is not fully
compatible with the propagation models (including the one we adopted here) which predict a monotonic CR gradient.

Even though the present analysis includes a dust template to account for the abundant missing gas 
present locally at the interface 
between the \HI\ and CO-bright phases, 
an alternative way to reconcile the flat emissivity profile
and a marked decline in CR density in the outer Galaxy is to invoke an increase in missing gas mass
with Galactocentric distance in the low metallicity environments of the outer Galaxy \citep[see
e.g.][]{Papadopoulos2002,Wolfire2010} beyond the local correction applied here.
We note that the large masses of dark gas in the outer Galaxy suggested by \citet{Papadopoulos2002}
(outweighing that of \HI\ by a factor of 5-15) might explain our results, 
whereas the remarkably constant dark gas fraction of ~30\% with mild dependence on metallicity 
suggested by \citet{Wolfire2010} is not sufficient 
to explain the large \HI\ emissivities measured by the LAT beyond the solar circle.

\section{Summary and Conclusion}

We have studied the diffuse $\gamma$-ray emission in the third Galactic quadrant using the first 18 months
of \textit{Fermi} LAT science data. Thanks to the excellent performance of the LAT,
we have obtained high-quality emissivity spectra of the atomic and molecular gas (traced by $W_{\rm CO}$)
in the 100~MeV -- 25.6~GeV energy range.

At the level of accuracy allowed by the LAT, the study of CR densities from $\gamma$-ray observations is now mostly limited by the
understanding of the ISM mass tracers, notably by the uncertainties in the derivation of atomic gas column
densities from \HI\ surveys and by the distribution of gas not accounted for by radio and microwave line surveys. 
In spite of those uncertainties, robust conclusions can be drawn 
concerning the ISM and CRs.

The molecular mass calibration ratio of the Local arm is found to be
$\sim 2 \times 10^{20}~{\rm cm^{-2}~(K~km~s^{-1})^{-1}}$,
significantly larger than that for the very local
Gould-Belt clouds in the second Galactic quadrant reported
by \citet{Abdo2010_2nd}. No significant differences of the ratio
are found between the Local arm and the interarm region.

No significant variations in the CR spectra are found across the outer Galaxy in the region studied,
and no large contrast in emissivity is seen in the interarm region between the Local and Perseus
arms (a contrast $<$
10--20\% is allowed by data). The measured gradient is much flatter than predictions
by a widely-used propagation model assuming that the CR source distribution largely peaks in the inner Galaxy.
A larger halo size and/or a flatter CR source distribution beyond the solar circle than those
usually assumed are required to reproduce the LAT data, while other scenarios such as a non-uniform diffusion coefficient 
or vast amounts of missing gas in the outer Galaxy are also possible.
Reliable determinations of the amount of atomic hydrogen in the plane are key 
to better constraining the property of CRs in our Galaxy.

\acknowledgments

The \textit{Fermi} LAT Collaboration acknowledges generous ongoing support
from a number of agencies and institutes that have supported both the
development and the operation of the LAT as well as scientific data analysis.
These include the National Aeronautics and Space Administration and the
Department of Energy in the United States, the Commissariat \`a l'Energie Atomique
and the Centre National de la Recherche Scientifique / Institut National de Physique
Nucl\'eaire et de Physique des Particules in France, the Agenzia Spaziale Italiana
and the Istituto Nazionale di Fisica Nucleare in Italy, the Ministry of Education,
Culture, Sports, Science and Technology (MEXT), High Energy Accelerator Research
Organization (KEK) and Japan Aerospace Exploration Agency (JAXA) in Japan, and
the K.~A.~Wallenberg Foundation, the Swedish Research Council and the
Swedish National Space Board in Sweden.

Additional support for science analysis during the operations phase is gratefully
acknowledged from the Istituto Nazionale di Astrofisica in Italy and the Centre National d'\'Etudes Spatiales in France.

This paper makes use of a development version of GALPROP provided by the GALPROP
team to the LAT collaboration solely for interpretation
of the LAT data.
GALPROP development is supported by NASA Grant NNX09AC15G.

We thank T. M. Dame for providing the moment-masked CO data.




\clearpage
\onecolumn

\clearpage

\setlength{\oddsidemargin}{-1cm}
\setlength{\evensidemargin}{-1cm}

\begin{deluxetable}{ccccccccccc}
\tabletypesize{\tiny}
\tablecaption{
A summary of fit parameters with 1 sigma statistical errors,
under the assumption of $T_{\rm S}=125~{\rm K}$. \label{tab:125K}
}
\tablehead{\colhead{Energy} 
& \colhead{$E^{2} \cdot q_{\rm HI,1}$} 
& \colhead{$E^{2} \cdot q_{\rm HI,2}$} 
& \colhead{$E^{2} \cdot q_{\rm HI,3}$} 
& \colhead{$E^{2} \cdot q_{\rm HI,4}$}
& \colhead{$E^{2} \cdot q_{\rm CO,1}$}
& \colhead{$E^{2} \cdot q_{\rm CO,2}$} 
& \colhead{${E^{2} \cdot q_{\rm CO,3}}^{\rm c}$} 
& \colhead{$E^{2} \cdot {q_{\rm EBV_{pos}}}$} 
& \colhead{$E^{2} \cdot {q_{\rm EBV_{neg}}}$} 
& \colhead{$E^{2} \cdot {q_{\rm iso}}$} \\
\colhead{GeV}& \colhead{} & 
\colhead{} & \colhead{} & \colhead{} & \colhead{} & \colhead{} & \colhead{} & \colhead{}
& \colhead{} & \colhead{}}
\startdata
0.10--0.14 & $1.19 \pm 0.15$ & $1.01 \pm 0.18$ & $0.88 \pm 0.22$ & $1.1 \pm 1.1$ & $0.0 \pm 0.1$ & $  7 \pm 6  $ & $\ 6 \pm 24 $ & $0.00 \pm 0.08$ & $1.04 \pm 0.45$ & $1.73 \pm 0.14$ \\
0.14--0.20 & $1.43 \pm 0.11$ & $1.23 \pm 0.13$ & $1.14 \pm 0.14$ & $0.7 \pm 0.8$ & $4.3 \pm 1.8$ & $5.0 \pm 2.9$ & $  0 \pm 1  $ & $0.77 \pm 0.36$ & $0.35 \pm 0.34$ & $2.01 \pm 0.09$ \\
0.20--0.28 & $1.60 \pm 0.08$ & $1.36 \pm 0.10$ & $1.25 \pm 0.11$ & $1.5 \pm 0.6$ & $5.0 \pm 1.3$ & $2.7 \pm 2.0$ & $  0 \pm 1  $ & $1.13 \pm 0.26$ & $0.32 \pm 0.24$ & $1.94 \pm 0.07$ \\
0.28--0.40 & $1.79 \pm 0.07$ & $1.57 \pm 0.08$ & $1.25 \pm 0.10$ & $1.7 \pm 0.5$ & $8.6 \pm 1.2$ & $5.2 \pm 1.6$ & $ 11 \pm 7\ $ & $1.28 \pm 0.21$ & $0.45 \pm 0.20$ & $1.90 \pm 0.06$ \\
0.40--0.56 & $1.81 \pm 0.07$ & $1.63 \pm 0.08$ & $1.61 \pm 0.10$ & $1.6 \pm 0.5$ & $7.5 \pm 1.1$ & $4.9 \pm 1.4$ & $  8 \pm 6  $ & $1.30 \pm 0.17$ & $0.87 \pm 0.18$ & $1.74 \pm 0.06$ \\
0.56--0.80 & $1.91 \pm 0.07$ & $1.62 \pm 0.07$ & $1.50 \pm 0.09$ & $2.0 \pm 0.5$ & $6.7 \pm 1.0$ & $4.7 \pm 1.3$ & $  3 \pm 5  $ & $1.43 \pm 0.17$ & $0.88 \pm 0.17$ & $1.49 \pm 0.06$ \\
0.80--1.13 & $1.75 \pm 0.07$ & $1.54 \pm 0.07$ & $1.48 \pm 0.09$ & $2.3 \pm 0.5$ & $8.7 \pm 1.0$ & $4.6 \pm 1.3$ & $ 16 \pm 5\ $ & $1.07 \pm 0.16$ & $0.88 \pm 0.16$ & $1.48 \pm 0.06$ \\
1.13--1.60 & $1.64 \pm 0.07$ & $1.45 \pm 0.08$ & $1.42 \pm 0.09$ & $1.4 \pm 0.5$ & $6.0 \pm 1.0$ & $3.7 \pm 1.3$ & $ 11 \pm 5\ $ & $1.13 \pm 0.16$ & $0.76 \pm 0.16$ & $1.28 \pm 0.06$ \\
1.60--2.26 & $1.60 \pm 0.08$ & $1.24 \pm 0.08$ & $1.07 \pm 0.09$ & $1.9 \pm 0.5$ & $6.0 \pm 1.0$ & $4.1 \pm 1.3$ & $  7 \pm 4  $ & $0.97 \pm 0.16$ & $0.53 \pm 0.16$ & $0.91 \pm 0.07$ \\
2.26--3.20 & $1.26 \pm 0.08$ & $1.02 \pm 0.08$ & $0.94 \pm 0.09$ & $1.3 \pm 0.5$ & $3.2 \pm 1.0$ & $3.1 \pm 1.3$ & $  3 \pm 4  $ & $0.82 \pm 0.16$ & $0.58 \pm 0.16$ & $0.85 \pm 0.07$ \\
3.20--4.53 & $0.80 \pm 0.09$ & $0.76 \pm 0.08$ & $0.75 \pm 0.09$ & $0.4 \pm 0.5$ & $5.7 \pm 1.0$ & $3.5 \pm 1.3$ & $ 11 \pm 4\ $ & $0.74 \pm 0.16$ & $0.42 \pm 0.16$ & $0.93 \pm 0.08$ \\
4.53--6.40 & $0.59 \pm 0.09$ & $0.41 \pm 0.08$ & $0.57 \pm 0.09$ & $0.6 \pm 0.5$ & $2.9 \pm 1.0$ & $3.6 \pm 1.3$ & $  0 \pm 0  $ & $0.74 \pm 0.16$ & $0.02 \pm 0.17$ & $0.86 \pm 0.08$ \\
6.40--9.05 & $0.51 \pm 0.09$ & $0.31 \pm 0.08$ & $0.28 \pm 0.09$ & $0.9 \pm 0.5$ & $2.2 \pm 1.0$ & $1.0 \pm 1.1$ & $  2 \pm 3  $ & $0.52 \pm 0.15$ & $0.07 \pm 0.16$ & $0.62 \pm 0.08$ \\
9.05--25.6 & $0.34 \pm 0.05$ & $0.22 \pm 0.05$ & $0.09 \pm 0.05$ & $0.7 \pm 0.3$ & $2.2 \pm 0.6$ & $1.4 \pm 0.7$ & $  3 \pm 2  $ & $0.07 \pm 0.06$ & $0.06 \pm 0.08$ & $0.49 \pm 0.05$ \\
\enddata
\tablenotetext{a}{
Units: $E^{2} \cdot q_{\rm HI, i} ({\rm 10^{-24}~MeV^{2}~s^{-1}~sr^{-1}~MeV^{-1}})$,
$E^{2} \cdot q_{\rm CO, i} ({\rm 10^{-4}~MeV^{2}~s^{-1}~cm^{-2}~sr^{-1}~MeV^{-1}~(K~km~s^{-1})^{-1}})$,
$E^{2} \cdot q_{\rm EBV} ({\rm 10^{-2}~MeV^{2}~s^{-1}~cm^{-2}~sr^{-1}~MeV^{-1}~mag^{-1}})$,
$E^{2} \cdot q_{\rm iso} ({\rm 10^{-3}~MeV^{2}~s^{-1}~cm^{-2}~sr^{-1}~MeV^{-1}})$
}
\tablenotetext{b}{The subscripts refer to four regions defined to perform the
analysis; 
1) Local arm, 2) interarm region, 3) Perseus arm, 4) beyond the
Perseus arm.}
\tablenotetext{c}{Some parameters are not well determined and their best-fit value is consistent
with 0. We present them for completeness.}

\end{deluxetable}

\begin{deluxetable}{ccccccccccc}
\tabletypesize{\tiny}
\tablecaption{
A summary of fit parameters with 1 sigma statistical errors, under the assumption of $T_{\rm S}=250~{\rm K}$.
\label{tab:250K}}
\tablehead{\colhead{Energy} 
& \colhead{$E^{2} \cdot q_{\rm HI,1}$} 
& \colhead{$E^{2} \cdot q_{\rm HI,2}$} 
& \colhead{$E^{2} \cdot q_{\rm HI,3}$} 
& \colhead{$E^{2} \cdot q_{\rm HI,4}$}
& \colhead{$E^{2} \cdot q_{\rm CO,1}$}
& \colhead{$E^{2} \cdot q_{\rm CO,2}$} 
& \colhead{${E^{2} \cdot q_{\rm CO,3}}^{\rm c}$} 
& \colhead{$E^{2} \cdot {q_{\rm EBV_{pos}}}$} 
& \colhead{$E^{2} \cdot {q_{\rm EBV_{neg}}}$} 
& \colhead{$E^{2} \cdot {q_{\rm iso}}$} \\
\colhead{GeV}& \colhead{} & 
\colhead{} & \colhead{} & \colhead{} & \colhead{} & \colhead{} & \colhead{} & \colhead{}
& \colhead{} & \colhead{}}
\startdata
0.10--0.14 & $1.35 \pm 0.07$ & $1.09 \pm 0.11$ & $1.20 \pm 0.11$ & $1.0 \pm 0.7$ & $0.1 \pm 0.7$ & $  9 \pm 6  $ & $\ 7 \pm 66 $ & $0.00 \pm 0.03$ & $0.96 \pm 0.18$ & $1.68 \pm 0.07$ \\
0.14--0.20 & $1.56 \pm 0.13$ & $1.33 \pm 0.15$ & $1.55 \pm 0.17$ & $0.5 \pm 0.8$ & $5.2 \pm 1.8$ & $7.1 \pm 3.0$ & $  0 \pm 1  $ & $0.79 \pm 0.33$ & $0.44 \pm 0.29$ & $1.97 \pm 0.10$ \\
0.20--0.28 & $1.82 \pm 0.10$ & $1.37 \pm 0.11$ & $1.66 \pm 0.13$ & $1.2 \pm 0.6$ & $6.2 \pm 1.3$ & $6.1 \pm 2.0$ & $  0 \pm 1  $ & $1.11 \pm 0.23$ & $0.24 \pm 0.20$ & $1.83 \pm 0.07$ \\
0.28--0.40 & $2.00 \pm 0.09$ & $1.70 \pm 0.09$ & $1.69 \pm 0.12$ & $1.7 \pm 0.6$ & $9.7 \pm 1.2$ & $7.5 \pm 1.6$ & $ 13 \pm 8\ $ & $1.19 \pm 0.19$ & $0.55 \pm 0.17$ & $1.83 \pm 0.06$ \\
0.40--0.56 & $1.95 \pm 0.08$ & $1.76 \pm 0.09$ & $2.11 \pm 0.11$ & $1.5 \pm 0.5$ & $8.4 \pm 1.1$ & $7.0 \pm 1.4$ & $  9 \pm 6  $ & $1.33 \pm 0.14$ & $0.83 \pm 0.14$ & $1.68 \pm 0.06$ \\
0.56--0.80 & $2.10 \pm 0.08$ & $1.76 \pm 0.08$ & $2.01 \pm 0.11$ & $2.1 \pm 0.5$ & $7.6 \pm 1.0$ & $6.4 \pm 1.3$ & $  4 \pm 5  $ & $1.33 \pm 0.15$ & $0.92 \pm 0.14$ & $1.42 \pm 0.06$ \\
0.80--1.13 & $1.90 \pm 0.08$ & $1.70 \pm 0.08$ & $1.95 \pm 0.10$ & $2.4 \pm 0.5$ & $9.6 \pm 1.0$ & $6.1 \pm 1.3$ & $ 17 \pm 5\ $ & $1.03 \pm 0.14$ & $0.88 \pm 0.14$ & $1.41 \pm 0.06$ \\
1.13--1.60 & $1.79 \pm 0.09$ & $1.59 \pm 0.09$ & $1.90 \pm 0.10$ & $1.4 \pm 0.5$ & $6.7 \pm 1.0$ & $5.0 \pm 1.3$ & $ 11 \pm 5\ $ & $1.05 \pm 0.14$ & $0.79 \pm 0.14$ & $1.23 \pm 0.07$ \\
1.60--2.26 & $1.74 \pm 0.09$ & $1.36 \pm 0.09$ & $1.45 \pm 0.11$ & $2.0 \pm 0.5$ & $6.6 \pm 1.0$ & $5.0 \pm 1.3$ & $  7 \pm 4  $ & $0.98 \pm 0.15$ & $0.57 \pm 0.14$ & $0.85 \pm 0.07$ \\
2.26--3.20 & $1.37 \pm 0.10$ & $1.13 \pm 0.09$ & $1.25 \pm 0.11$ & $1.5 \pm 0.5$ & $3.7 \pm 1.0$ & $3.8 \pm 1.3$ & $  3 \pm 4  $ & $0.80 \pm 0.14$ & $0.57 \pm 0.14$ & $0.79 \pm 0.08$ \\
3.20--4.53 & $0.84 \pm 0.10$ & $0.85 \pm 0.09$ & $0.99 \pm 0.11$ & $0.4 \pm 0.5$ & $6.1 \pm 1.0$ & $4.0 \pm 1.3$ & $ 11 \pm 4\ $ & $0.71 \pm 0.14$ & $0.43 \pm 0.14$ & $0.91 \pm 0.08$ \\
4.53--6.40 & $0.65 \pm 0.10$ & $0.48 \pm 0.09$ & $0.74 \pm 0.11$ & $0.7 \pm 0.5$ & $3.2 \pm 1.0$ & $3.9 \pm 1.3$ & $  0 \pm 0  $ & $0.65 \pm 0.15$ & $0.13 \pm 0.15$ & $0.84 \pm 0.09$ \\
6.40--9.05 & $0.54 \pm 0.10$ & $0.34 \pm 0.09$ & $0.37 \pm 0.10$ & $0.9 \pm 0.5$ & $2.5 \pm 1.0$ & $1.1 \pm 1.1$ & $  2 \pm 3  $ & $0.55 \pm 0.14$ & $0.06 \pm 0.15$ & $0.59 \pm 0.09$ \\
9.05--25.6 & $0.38 \pm 0.08$ & $0.23 \pm 0.13$ & $0.13 \pm 0.10$ & $0.7 \pm 0.4$ & $2.3 \pm 0.6$ & $1.5 \pm 0.7$ & $  3 \pm 2  $ & $0.1 \pm 0.3$ & $4 \pm 1$ & $0.45 \pm 0.14$ \\
\enddata
\tablenotetext{a}{
Units: $E^{2} \cdot q_{\rm HI, i} ({\rm 10^{-24}~MeV^{2}~s^{-1}~sr^{-1}~MeV^{-1}})$,
$E^{2} \cdot q_{\rm CO, i} ({\rm 10^{-4}~MeV^{2}~s^{-1}~cm^{-2}~sr^{-1}~MeV^{-1}~(K~km~s^{-1})^{-1}})$,
$E^{2} \cdot q_{\rm EBV} ({\rm 10^{-2}~MeV^{2}~s^{-1}~cm^{-2}~sr^{-1}~MeV^{-1}~mag^{-1}})$,
$E^{2} \cdot q_{\rm iso} ({\rm 10^{-3}~MeV^{2}~s^{-1}~cm^{-2}~sr^{-1}~MeV^{-1}})$
}
\tablenotetext{b}{The subscripts refer to four regions defined to perform the
analysis; 
1) Local arm, 2) interarm region, 3) Perseus arm, 4) beyond the
Perseus arm.}
\tablenotetext{c}{Some of parameters are consistent with 0 and thus
are not well determined. We present them for reference.}
\end{deluxetable}

\begin{deluxetable}{cc}
\tabletypesize{\scriptsize}
\tablecaption{
Number of counts in each energy bin.
\label{tab:counts}}
\tablehead{\colhead{Energy (GeV)} & \colhead{Counts}}
\startdata
0.10-0.14 & 26673 \\
0.14-0.20 & 71637 \\
0.20-0.28 & 91336 \\ 
0.28-0.40 & 93286 \\ 
0.40-0.56 & 78330 \\
0.56-0.80 & 61337 \\
0.80-1.13 & 45386 \\
1.13-1.60 & 30713 \\
1.60-2.26 & 19351 \\
2.26-3.20 & 11301 \\
3.20-4.53 & 6426 \\
4.53-6.40 & 3761 \\
6.40-9.05 & 2095 \\
9.05-25.6 & 2333 \\
\enddata
\end{deluxetable}

\begin{deluxetable}{ccccc}
\tabletypesize{\scriptsize}
\tablecaption{
Log-likelihood and emissivities for several choices of $T_{\rm S}$.
\label{tab:lnL}
}
\tablehead{\colhead{$T_{\rm S}$} 
& \colhead{$\ln(L)$}
& \colhead{$q_{\rm HI,1} (E \ge {\rm 100~MeV})$} 
& \colhead{$q_{\rm HI,2} (E \ge {\rm 100~MeV})$} 
& \colhead{$q_{\rm HI,3} (E \ge {\rm 100~MeV})$} }
\startdata
100 K & 114407.6 & $1.32 \pm 0.04$ & $1.27 \pm 0.05$ & $0.86 \pm 0.06$ \\
125 K & 114480.1 & $1.47 \pm 0.05$ & $1.26 \pm 0.06$ & $1.14 \pm 0.08$ \\
250 K & 114533.8 & $1.62 \pm 0.04$ & $1.35 \pm 0.05$ & $1.53 \pm 0.06$ \\
400 K & 114544.5 & $1.67 \pm 0.07$ & $1.39 \pm 0.08$ & $1.64 \pm 0.09$ \\
optically-thin & 114552.8 & $1.70 \pm 0.07$ & $1.39 \pm 0.07$ & $1.77 \pm 0.09$ \\
\enddata
\tablenotetext{a}{
Units: $q_{\rm HI, i} ({\rm 10^{-26}~photons~s^{-1}~sr^{-1}~H\mathchar`-atom^{-1}})$}
\tablenotetext{b}{The subscripts refer to four regions defined to perform the
analysis; 
1) Local arm, 2) interarm region, 3) Perseus arm.
}
\end{deluxetable}

\clearpage

\end{document}